\documentclass[10pt]{article}

\newcommand{\be}{\begin{equation}}
\newcommand{\ee}{\end{equation}}
\newcommand{\ba}{\begin{eqnarray}}
\newcommand{\ea}{\end{eqnarray}}

\begin{document}
\sloppy

\begin{center}
\centering{\LARGE \bf Earthquakes\,: from chemical alteration to mechanical rupture}
\end{center}
\vskip 0.5cm
\begin{center}
\centering{\bf Didier Sornette \footnote{email\,: sornette@cyclop.ess.ucla.edu}}
\end{center}
 
\begin{center}
\centering{\bf Department of Earth and Space Science\\ and Institute of Geophysics
and Planetary Physics\\ University of California, Los Angeles, California 90095\\
and\\
Laboratoire de Physique de la Mati\`ere Condens\'ee\\ 
CNRS and Universit\'e des
Sciences, B.P. 70, Parc Valrose\\ 06108 Nice Cedex 2, France\\
}
\end{center}

\begin{center}
\centering{\today}
\end{center}

\vskip 1cm

{\large\bf Abstract}\,: In the standard rebound theory of earthquakes,
elastic deformation energy is progressively stored in the crust until a threshold
is reached at which it is suddenly released in an earthquake. We review three
important paradoxes, the strain paradox, the stress
paradox and the heat flow paradox, 
that are difficult to account for in this picture, either
individually or when taken together. Resolutions of these
paradoxes usually call for additional assumptions on the nature of
the rupture process (such as novel modes of deformations and ruptures)
prior to and/or during an earthquake, on the nature of the fault and on the effect of trapped
fluids within the crust at seismogenic depths. We review the evidence for the
essential importance of water and its interaction with the modes of deformations.
Water is usually seen to have mainly the mechanical effect of decreasing the normal
lithostatic stress in the fault core
 on one hand and to weaken rock materials via hydrolytic weakening
and stress corrosion on the other hand. We also review the evidences that 
water plays a major role in the alteration of minerals 
subjected to finite strains into other structures in out-of-equilibrium conditions. 
This suggests novel exciting routes to understand what is an earthquake, that
requires to develop a truly multidisciplinary approach involving mineral chemistry,
geology, rupture mechanics and statistical physics.

\pagebreak

\tableofcontents

\pagebreak

\section{Introduction}

A seismological map of the Earth reveals that the epicenters
of earthquakes are for the most part located on long arcing strips - the
seismological belts that encircle the planet. These seismological belts
mark the Plate Tectonic boundaries of convergence, divergence and transform
that define the lithosphere plates of the Earth. Much of the active
seismological belts lie close to or are underneath densely populated areas
of human habitation. 

The fact that large populations are often found in tectonically active,
earthquake prone areas may be due to the favorable environment they provided for
the development of prehistoric
human populations [{\it Baily et al.}, 1993; {\it King et al.}, 1994].
Furthermore, an important factor in the development of ancient
civilizations [{\it Knopoff and Sornette}, 1995], 
i.e. the development of large communal populations,
may have been the geographical isolation, induced by tectonic processes
of mountain building that presented natural defenses. Examples of such
ancient  civilizations are to be found in Anatolia, 
China, Greece, India, Japan, Java, Mexico, 
Persia, Peru and Rome. The absence of tectonic defences in the
plains promoted diffusion by marauders and hence the incubation processes
necessary to develop large populations could not emerge.
The growth of these populations has continued to this day.
It was much later, in the middle ages, that global trade emerged as
an even more potent force that allowed for the development of large
populations in the plains, for example, of northern and western Europe.
And the exploration and development of North America took place because
of these more potent forces as well. An additional factor is that 
earthquake prone areas are geologically active and thus rich in mineral and oil
reservoirs, as well as in the quality of the soil for agriculture exploitation.
It has been estimated that the economic balance between property losses incurred during 
earthquakes and gains from their indirect economic benefits is largely positive.

Thus, people continue to live in large metropolitan areas 
such as Tokyo, San Francisco, Los Angeles,
Santiago, and Rome to name a few, which are in danger of destruction by
earthquakes. Statistics compiled by insurance companies such as Lloyds of
London, rate earthquakes first among natural disasters in terms of severity
of damage and loss of life. The last destructive japanese earthquake at Kobe of 
magnitude 6.9 that occurred on January, 17, 1995, led to about $\$~200$ billions
of damage in a country that is a leader in engineering construction safety rules and hazard 
reduction efforts. Estimates by Japanese banks and insurances put the damage for the next
great Tokyo earthquake \footnote{The last great Tokyo earthquake occurred in 1923 and
completely destroyed the city. The estimate recurrence time is about $70$ years plus or
minor a large uncertainty.} at more than a trillion dollar, with important economic
repercussions worldwide in the following years. Recent careful analysis of the seismic
energy-frequency Gutenberg-Richter relation using an independent regionalization
of global seismicity, the Harvard centroid-moment tensor catalog data and a 
combination of observed rates and tectonic deformation rates shows that 
all regions, except mid-ocean ridges, are susceptible to the same
maximum earthquake magnitude around $8.5$ [{\it Kagan}, 1997].
According to this result \footnote{This view is controversial in comparison
to the standard way of estimating the largest possible earthquake, based on the maximum
fault length, which thus leads in general to much smaller estimations.
 However, the limitations of this 
 ``old'' approach are more and more recognized because most recent damaging earthquakes 
 occurred on unknown faults or activated several disconnected faults.}, there are
almost no regions that are safe from great earthquakes on the long term. The only difference is the 
typical time scales, tens of years in Chile, a hundred or a few hundreds of years in Japan
and California to tens of thousands of years in so-called non-seismic regions. 
These time scales are simply controlled by the deformation rates, i.e. the 
rate with which faults are reloaded, which go from several centimeters per year
to a fraction of millimeter per year in different parts of the world.
These time scales separating major earthquakes
are larger than the historical development time of 
our industrial society leading to a global interconnected
economy. We can thus expect surprising impacts that have
never been experienced in the past.

Since the inception of seismology, there
has been a great urgency upon the search for reliable precursor phenomena
of earthquakes and their generative processes [{\it Gokhberg et al.}, 1995].
It is widely accepted that the extensive efforts of the last thirty years
to find reliable earthquake prediction methods have largely failed
[{\it Evans}, 1997; {\it Geller et al.}, 1997; {\it Scholz}, 1997; 
{\it Snieder and van Eck}, 1997; {\it Wyss et al.}, 1997]. This has
even led some to assert that earthquakes are unpredictable [{\it Geller et al.}, 1997].
More researchers agree now that it is necessary to improve our fundamental 
understanding because ``black box'' prediction schemes do not work reliably.
This review attempts to bring up to the physicists a flavor of 
the crisis [{\it Jordan}, 1997; {\it Kagan}, 1998] and 
the challenges felt by the earthquake science community.

\section{Earthquakes\,: where are the challenges?}

\subsection{The $N$-body problem}

The most common view among physicists interested in earthquake modelling is that
the real challenge lies in its ``N-body'' nature. This has led to a flurry of
activities starting in the late eighties with hundreds of publications on earthquakes 
in physical as well as geophysical journals since.
To simplify to the extreme, the one-earthquake or one-fault problem is usually thought to be 
relatively well-understood (even if an ``isolated'' fault has never been observed!)
and the excitement (for the physicists) 
emerges when coupling (via long-range elasticity)
many faults presenting highly nonlinear responses (threshold dynamics) in the presence of
quenched (and also annealed) heterogeneity (to account for inhomogeneous rocks and fault
geometries). 

\subsection{Self-organized criticality}

This is the vision that led to the proposed analogy between earthquake organization
and self-organized criticality [{\it Sornette and Sornette}, 1989; {\it Bak and Tang}, 1989; 
{\it Sornette et al.}, 1990; {\it Ito and Matsuzaki}, 1990;
{\it Scholz}, 1991; {\it Sornette}, 1991] (see [{\it Main}, 1996; {\it Grasso and
Sornette}, 1998] for reviews) and 
the recognition that earthquake dynamics offer for the solid earth a complexity and richness
(and difficulties!) equivalent or even greater to 
that of hydrodynamic turbulence [{\it Kagan}, 1994]. Self-organized
criticality provides a conceptual framework for
understanding the Gutenberg-Richter size distribution of earthquakes, 
the power law fault length distribution, the fractal geometry
of sets of earthquake epicenters and of fault patterns among others [{\it Sornette}, 1991]. 
It has recently been exploited to advance our understanding on the crust organization and about the
rich and subtle properties found in tectonics and seismology, such as
induced seismicity by water impoundment, mineral and gas extraction [{\it Grasso and Sornette}, 1998].

In a nutshell, SOC refers to
the spontaneous organization of a system driven from outside in a globally dynamical
statistical stationary state, which is characterized by self-similar
distributions of event sizes and fractal geometrical properties.
SOC applies to the class of phenomena occurring in continuously
driven out-of-equilibrium systems made of many interactive components, which
possess the following fundamental properties : 1) a highly non-linear behavior,
2) a very slow driving rate , 3) a globally stationary  regime, 
characterized by stationary statistical properties,
and 4) power distributions of event sizes and fractal geometrical properties.
The crust obeys these four conditions [{\it Sornette}, 1991]. 
\begin{enumerate}
\item The threshold response can be associated with the 
rupture instability leading to the sudden unlocking of the fault.

\item The slow driving rate is that of the slow tectonic
deformations thought to be exerted at the borders of a given tectonic plate by
the neighboring plates and at its base by the underlying lower crust and mantle.
The large separation of time scales between the driving tectonic velocity
$(~cm/year)$ and the velocity of slip  $(~m/s)$ makes the crust problem maybe the
best natural example of self-organized criticality, in this sense. 

\item The stationarity condition ensures that the system is not in a
transient phase, and distinguished the long-term organization of faulting in the
crust from, for instance, irreversible rupture of a sample in the laboratory. 

\item The power laws and fractal properties reflect the notion of scale invariance,
namely measurements at one scale are related to measurements at another scale by
a normalization involving a power of the ratio of the two scales.  These
properties are important and interesting because they characterize systems with
many relevant scales and long-range interactions as probably exist in the crust.
\end{enumerate}

A remarkable feature of the SOC state is in the way it emerges. The spatial
correlations between different parts of the system do not appear as a result of a
progressive diffusion from a nucleus but from the repetitive action of rupture
earthquake cascades. In other words, within the SOC hypothesis,
different portions of the crust become correlated at long distances by the action
of earthquakes and their relaxations 
which ''transport'' the fluctuations of the alteration and stress fields in the different
parts of the crust many times back and forth to finally organize the system. 
This physical picture is substantiated by various numerical and analytical
studies of simplified models of the crust [{\it Bak and Tang}, 1989; {\it Zhang}, 1989; 
{\it Chen et al.}, 1991; {\it Cowie et al.}, 1994]. 

The crust is heterogeneous at many scales from the smaller atomic scale, with 
the presence of dislocations, impurities and grain boundaries to the scale of
tectonic plate boundaries. Its structure can be approximatively described as
fractal, probably better by multifractal or even better as hierarchical [{\it Ouillon et al.},
1996]. Water is probably found everywhere within the crust, however with
a very large heterogeneity. We thus envision the crust containing a 3D hierarchical network of
faults and discontinuities with water filling them partially in varying amounts.
As a consequence and in the presence of the heterogeneous strain field, rock
alteration varies from place to place. Integrated over time, a
substantial fraction of the
crust is close to rupture instability. However, at a given instant of time, only 
a small subset of the crust is altered, sufficiently stressed  and riped 
enough to destabilize into a rupture earthquake.
Together with the localization of
seismicity on faults, this leads to the conclusion that a significant fraction of
the crust is susceptible to rupture, while presently being quiescient.  The
quantitative determination of the susceptible fraction is dependent on the
specificity of the model [{\it Zhang}, 1989; {\it Pietronero et al.}, 1991; 
{\it Sornette et al.}, 1994] and cannot thus be ascertained with precision for the crust. What is
important however in that the susceptible part of the crust can be activated with
relatively small perturbations or by modification of the overall driving
conditions. This remark leads to a straighforward interpretation of induced
seismicity by human activity [{\it Grasso and Sornette}, 1998].

\subsection{Application of statistical physics concepts}

Other modern physical concepts has been shown to be relevant to earthquakes. For instance, 
faults result from growth
processes [{\it Sornette et al.}, 1990] and, in some limited
 cases, converge to geometrical structures
that can be mapped onto optimal random manifolds [{\it Miltenberger et al.}, 1993; {\it Sornette 
et al.}, 1994a].
Nucleation and spinodal decomposition has
been suggested to describe an incipient earthquake [{\it Rundle and Klein}, 1995; {\it Klein 
et al.}, 1997; {\it
Dieterich}, 1992].
The presence of quenched heterogeneity coupled with dynamics is also thought to play a
fundamental role [{\it Miltenberger et al.}, 1993; {\it Sornette et al.}, 1994b;
{\it Knopoff}, 1996a; {\it Fisher et al.}, 1997] and leads to complex spatio-temporal
behavior. Long-range elasticity can induce ``frustration'' in the motion of adjacent
tectonic blocks [{\it Sornette et al.}, 1994a; {\it Sornette and Vanneste}, 1996]
and lead to a hierarchical organization of
faults and earthquakes in space and time [{\it Alekseevskaya et al.}, 1977; {\it Scholz}, 1990; {\it 
Ouillon et al.}, 1996], suggested to be
qualitatively similar to spinglasses [{\it Sornette et al.}, 1994b]. 
The tensorial nature of crust motions leads to 
kinematic compatibility and geometric compatibility requirements [{\it Gabrielov et al}, 1996].
Kinematic compatibility (also known as St-Venant compatibility) stems from the fact that the 
displacement must be uni-valued along a closed countour. Geometric compatibility
measures the tendancy for tectonic blocks to create or anihilate surface, by the 
activation of so-called thrust and normal faulting that relax the motion along the vertical
direction perpendicular to the tectonic plates [{\it Gabrielov et al}, 1996]. 

Recent efforts have been undergoing at the frontier between seismology and statistical
physics to understand the origin of the observed complexity of the spatio-temporal
patterns of earthquakes. An interesting debate has been whether 
space-time complexity can occur on an homogeneous fault solely from the nonlinear
dynamics [{\it Shaw}, 1993; 1995; 1997; {\it Cochard and Madariaga}, 1994; 1996]
associated to the slip and velocity dependent friction law
[{\it Dieterich and Kilgore}, 1994; {\it Dieterich}, 1992]. Or, is the presence
of quenched heterogeneity necessary [{\it Rice}, 1993; {\it Benzion and Rice}, 1993; 1995;
{\it Knopoff}, 1996b]? It is now understood that complexity can emerge purely 
from the nonlinear laws but
heterogeneity is probably the most important factor dominating the multi-scale
complex nature of earthquakes and faulting [{\it Ouillon et al.}, 1996]. It is also known to 
control the appearance of self-organized critical behavior in a class of models
relevant to the crust [{\it Sornette et al.}, 1995; {\it Shnirman and Blanter}, 1998].

\subsection{The one-body problem}

More than six years ago, a first draft of this review 
was focused on statistical models of rupture in
heterogeneous media and their self-organizing principles,
with emphasis on cooperative behavior in order to account for the 
many observed power law behaviors
(see for instance [{\it Kagan}, 1994; {\it Knopoff}, 1996a; 1997]). Carlson et al. 
[1994],  Rundle and Klein [1995] and Langer et al. [1996] have written reviews in this spirit, 
however mostly specialized to spring-block models.

In the last few years, my point of view has evolved dramatically. I am still convinced that the
dynamics of earthquakes and their space-time structure
provide a nice example of self-organizing nonlinear spatio-temporal
systems and that this deserves investigations. I am confident that the 
physical community does not need encouragement in this area.
However, I have come to realize that maybe the biggest challenge of all resides
in the so-called ``one-body'' problem, i.e. in the understanding of the single earthquake ``cycle''.
This came while I was trying to synthetize many geological observations and
experimental data. It progressively dawned on me that even the supposedly ``simple'' one-earthquake 
process is not at all understood and that this poses many fundamental questions.
This review is focused
at the basic facts that poses difficulties for their interpretation and for making
them compatible with the standard pictures. The review is thus organized around
the three main paradoxes, the strain, stress and heat flow paradoxes. 
The emphasis is on geological, seismological, mechanical and chemical data and concepts rather than 
on statistical physics. 
The challenge is to attempt to synthetize and account for the large body of phenomenology existing in 
these diverse disciplines that are rarely put together. The style of work is similar to that
of a detective slowly and patiently assembling clues to identify the ``shooting gun''\,:
what is really an earthquake?
It is hoped that this will help realize
that the one-earthquake problem is amazingly diverse and multidisciplinary in 
essence and that mechanical and chemical processes are intimately coupled in earthquake
processes. The main theme developed here is that it is not possible to get a reasonable 
description of an earthquake if one forgets about the chemical processes occurring at the
small scale. In contrast to the homogeneization approaches that lead to treatments of
earthquakes solely based on macroscopic descriptions, here we propose 
that the preparation of an earthquake and its triggering start at the atomic level and that
keeping such a (multi-scale) description is necessary for understanding what an earthquake is. 
This may be an example where different levels of the hierarchical structure of science
meet and interfere, in contrast to the more usual bottom-up structure [{\it Anderson}, 1972].

\section{The mechanical view}

The fault of the geologist is
not the same object as the fault of the rock mechanician which is itself a
different structure than the fault of the seismologist. For the geologist, complex
 cataclistic deformations are ubiquitous in fault zones. For the
rock mechanician, gouge \footnote{The gouge is the fragmented core of a fault.}
 and localized deformation are the key elements. For the
seismologist, the earthquake source as one (or several) double-couple set(s) of
forces (see section \ref{errrrrzz} for a definition of a double-couple)
on well-defined rupture planes is the fundamental quantity. We shall try 
to synthetize these different view points and, in this way, gain a
better understanding of the physical mechanism of earthquakes.

\subsection{Seismology versus rock mechanics}

It is a common wisdom that an earthquake is primarily a mechanical process which
appears as a genuire rupture (or a reactivation of a previous rupture) of the crust,
and the earth behaves as an elastic body during the short time span of the phenomena
[{\it Aki}, 1988]. Such model is very good at calculating the static strain fields
that are created by large dislocations \footnote{A dislocation is a planar fault domain
over which a slip discontinuity is imposed.} at fault planes. They are also very good at
calculating the propagating compressional and shear waves that are important
dynamical consequences of earthquakes. It is recognized that some complication
occurs in the earthquake source region before and during the earthquake but it is
encapsulated in the purely mechanical formulation in terms of the double-couple
\footnote{An admissible earthquake must be such that the medium is 
in mechanical equilibrium after the introduction of this event. Thus, the total
force and the total torque induced by the earthquake must vanish, hence the term 
``double-couple''.} source mechanism [{\it Burridge and Knopoff}, 1964]. At the simplest level of
description, elastic energy is progressively stored until a static friction
threshold is reached at which the fault slips according to dynamical friction,
creating the equivalent of an extended dislocation.

This description is evidenced by the many superficial observations of fresh surface
rupture traces, fresh scarps and fresh displacements seen at the surface of the
earth after large earthquakes. Fractures are also widely observed in surface rocks
of all types. When trying to understand the mechanism underlying
an earthquake, it thus seems natural to think that some sort of rupture is the {\it
cause} of the earthquake. The standard earthquake picture is thus based on the idea
that pre-existing cracks can grow progressively by subcritical growth, leading
ultimately to a catastrophic failure propagating at a fraction of the shear wave
speed. A similar and related mechanism is to invoke friction of pre-existing
faults, which goes from very slow stable growth to unstable, depending on the
stress loading and other water content.

The present understanding is vividly summarized by 
Scholz [1992a], who contrasts the two main paradigms of the earthquake mechanism.
On the one hand, seismologists tend to view an earthquake as a seismic source
characterizing a fracture phenomenon, {\it i.e.} the rupture of a strong part of
the crust. This view is influenced by the seismological observations which put the
emphasis on the generation of elastic waves and their propagation to the recording
stations. On the other hand, rock mechanicians
consider an earthquake as a stick-slip event controlled by the friction properties
of the fault, {\it i.e.} the destabilization of a weak part of the crust. 
This formulation has been shaped by numerous laboratory experiments under a
variety of pressure and temperature conditions (which however reproduce only
imperfectly the conditions prevailing in the crust). Scholz
[1992a] sees the transition from the first paradigm to the second one as
ineluctable\,: the seismological paradigm uses concepts such as
asperities and barriers which remain loosely defined (see however for instance 
[{\it Aki}, 1979])\,; in contrast, the stick-slip friction mechanism has developed
quantitatively both with laboratory experiments and theoretical formulations. 

\subsection{Friction}

As a scientific field, solid friction has a long history, probably starting in
the western world about 500
years ago with Leonardo di Vinci, continuing with the empirical Amontons'laws two centuries
later and Coulomb's investigations of the influence of slipping velocity on friction in the
XVIII century. Only three decades ago was it recognized that friction has probably a fundamental
role in the mechanics of earthquakes [{\it Brace and Byerlee}, 1966].
Rock mechanicians
consider an earthquake as a stick-slip event controlled by the friction properties
of the fault, i.e. the destabilization of a weak part of the crust. 
Numerous laboratory experiments have been carried out to identify the parameters that
control solid friction and its stick-slip behavior [{\it Persson and Tosatti}, 1996]. 
The most significant variables
appear to be the mineralogy, the porosity, the thickness of the gouge, the effective
pressure, the temperature and the water content [{\it Byerlee et al}, 1968;
{\it Brace}, 1972; {\it Beeman et al.}, 1988;
{\it Gu and Wong}, 1991; {\it Johansen et al.}, 1993; {\it Streit}, 1997]. Low velocity
experiments have established that solid friction is a function of both the velocity of 
sliding and of one or several state parameters, roughly quantifying the true surface of contact
[{\it Brace}, 1972; {\it Dieterich}, 1972; 1978; 1979; 1992; {\it Ruina}, 1983; {\it Cox}, 1990; 
{\it Beeler et al.}, 1994; 1996; {\it Baumberger and Gauthier}, 1996; {\it Scholz}, 1998].
These Ruina-Dieterich laws constitute the basic ingredients in most models and numerical
elastodynamic calculations that attempt to understand the characteristics of earthquake sources.

A well-known and serious limitation of these calculations based on
laboratory friction experiments is that the friction laws have been determined using
sliding velocities no more than about $1$~cm/s, i.e. 
orders of magnitude below the sliding velocity of
meters or tens of meters believed to occur during an earthquake. Is it correct then to 
extrapolate these laws, and especially the velocity weakening dependence, to large velocities?
This question is all the more relevant when one examines the underlying physical 
mechanisms of the friction laws. At law velocity, hysteretic elastic and plastic 
deformations at the scale of roughness asperities seem to play a dominant role
[{\it Bowden and Tabor}, 1954; {\it Sokoloff}, 1984; {\it Jensen et al.}, 1993; 
{\it Dieterich and Kilgore}, 1994; {\it Caroli and Nozi\`eres}, 1996;
{\it Tanguy and Nozi\`eres}, 1996; {\it Tanguy and
Roux}, 1997; {\it Caroli and Velicky}, 1997; {\it Bocquet and Jensen}, 1997]. 

At large velocities, different mechanisms come into play.
Collisions between asperities and transfer of momentum between
the directions parallel and perpendicular to the motion may become an important 
mechanism [{\it Lomnitz}, 1991; {\it Pisarenko and Mora}, 1994]. This regime
is probably relevant for the resolution of the apparent low-friction problem stemming
from the heat flow paradox and the low strength of the San Andreas fault as we 
discuss below. 
Recently, Tsutsumi and Shimamoto [1996; 1997] have reached a completely novel 
regime, by performing friction measurements
on rotating cylindrical samples at velocities up to $1.8$~m/s and for slips
of several tens of meters. They results cannot be taken completely for granted due
to several experimental problems but seem to indicate the existence of 
a change of regime from velocity weakening to velocity strengthening and then again
to velocity weakening at the largest velocities. This last regime seems to be
associated to the melting of a very thin layer. 
It is quite possible that different
physical mechanisms might lead to a change of regime in the velocity and slip
dependence of the solid friction law at large slip velocities. 
One thus needs to explore the high velocity
regime in a variety of conditions, as has been done for the low velocity regime.
In this goal, recent realistic 3D numerical simulations of elastic bodies 
sliding on top of each other in
a regime of velocities of meters to tens of meters per second have been performed
that allows one to probe more intimately the response of the bodies and the nature of the friction,
that any experiment could. A velocity
strengthening is found with a doubling of the friction coefficient when the velocity increases
from $1~m/s$ to $10~m/s$, which reflects the increasing strength of vibrational damping
[{\it Maveyraud et al.}, 1998].

\subsection{The standard model\,: elastic rebound \label{rebound}}

Geodetic measurements of crustal deformations starting in 1850 in California and
conducted after the 1906 San Francisco earthquake indicated that significant
horizontal displacements parallel to the ruptured San Andreas fault had occurred
both before and after the earthquake. In particular, Reid noticed that distant
points of opposite sides of the fault had moved $3.2$ meters over the 50-year
period prior to 1906. This led him to the famous {\it elastic rebound} theory for
the cause of earthquakes [{\it Bolt}, 1993].

The theory is best explained using a simple diagram shown in figure 1,
representing a locked fault within an elastic medium immediately after the last
earthquake [{\it Turcotte and Schubert}, 1982]. A relative uniform tectonic velocity $u_0$ is applied
at a distance $b$ from each side of the fault and the shear strain $\epsilon$
increases with time according to  
\be
\epsilon = {u_0 t \over 2b} ~.
\ee
The stress increases correspondingly uniformly as
\be
\sigma = \sigma_d + G {u_0 t \over 2b}  ~,
\label{stress}
\ee
where $\sigma_d$ is the so-called dynamical frictional stress that is operative
on the fault at the end of faulting and $G$ is the shear modulus.
The locked fault can transmit any shear stress less than its assumed static
frictional stress $\tau_f$. When this stress is reached at the time
\be
t^* = {2b \over G u_0} (\sigma_f - \sigma_d)~,
\ee
slip occurs on the fault in a few seconds so that the edges of the plates can be
assumed to be stationary during this time. The accumulated shear strain
$\epsilon^* = {u_0 t^* \over 2b}$ is recovered by the plates by a springing back or
rebounding on each side of the fracture, thus the term {\it elastic rebound}. The
resulting displacement on the fault is $d=2\epsilon^* 2 b$. After the earthquake,
the fault locks and the cycle repeats.

\section{The strain paradox}

From the observation of the fault slip $d$ of the earthquake, the elastic
rebound theory of the previous section with eq.(\ref{stress}) predicts the order of
magnitude of the width $b$ over which the shear strain develops progressively
across the fault prior to the earthquake\,: 
\be
2b = {G  \over (\sigma_f - \sigma_d)} d~.
\label{bbb}
\ee
We consider a typical displacement $d \approx 5 m$ on the fault during a large
earthquake. The shear elastic modulus is of the order of $10 ~GPa$ \footnote{Basalt
and granite have a large modulus $30-50~ GPa$. However, the crust is heterogeneous
with defects and cracks are many different scales. As a consequence, the modulus
determined in the laboratory from homogeneous samples overestimates significantly the
values in the field. Due to the large heterogeneity, it is difficult to give a
meaningful and precise value for the modulus. Our calculation must rather be taken as
an order-of-magnitude estimate.}. The
stress drop  $\sigma_f - \sigma_d$ on the fault during the large earthquake lie in
the range $1-10 MPa$. The equation (\ref{bbb}) leads to the estimate  
\be
2b = 5-50 ~km~.
\label{resultb}
\ee
Note that $b$ is all the smaller, the larger is the stress drop.

The order of magnitude of the result (\ref{resultb}) is robust with respect to the 
many ways with which one can refine the model. A standard model [{\it
Turcotte and Schubert}, 1982] (page 371) consists of two lithospheric plates of
thickness $T$ sliding past each other and in contact at the position of the fault.
Initially, there is not stress on the fault which is locked to a depth $a$. This
depth represents the thickness of the brittle seismogenic crust. Below,
the material is ductile and flows continuously without locking. The plates are
then displaced progressively but no displacement occurs on the locked part of the
fault. Assuming that the displacement rate is imposed at a large distance
from the fault, a simple elastic calculation [{\it Turcotte and Schubert}, 1982]
(page 373) shows that the shear strain is concentrated in a narrow zone of width
$b \approx a$ around the fault. For a typical thickness $a \approx 15 ~km$ for
the brittle crust, we thus get a confirmation of the order of magnitude estimate 
(\ref{resultb}).

Modern geodetic measurements are sufficiently precise to test for the existence
of strain localization. The surprise is that there is no geodetic evidence of the
existence of a concentration of shear deformation at the scale of $1-10~km$ 
around faults in many situations, even if others exhibit it [{\it Pearson et al.}, 
1995]. The geodetic displacement profiles obtained across section of
faults that have not ruptured in the last decades give an essentially uniform
strain over distances of $150 km$ or more from the fault [{\it Walcott et al.}, 1978;
1979; {\it Shen et al.},
1996; {\it Snay et al.}, 1996]. This leads to a problem. Indeed, if we put $2b
\approx 150 km$ in (\ref{stress}) while keeping the plate tectonic displacement
of the same order $d \approx 5 m$, we get a reduction of strain by a factor
$10-100$. This leads to a much smaller stress drop at rupture than usually
measured for large earthquakes. Taking  $G \approx 10 ~GPa$ leads to $\sigma_f -
\sigma_d \approx 0.1 MPa$ ($1$ bars) (see below the discussion in section
\ref{friction}). We note that there is evidence that some small earthquakes can
exhibit very small stress drop of the order of $0.1 MPa$ [{\it Scholz}, 1990]
(page 183,figure 4.10), but this is by no mean the case for the population of
large earthquakes. How a large earthquake can develop and pass over barriers with
a small stress drop is another unresolved difficult question in the standard
conceptual framework.

On the other hand, strain concentrations are observed very
neatly {\it after} a large earthquake [{\it Shen et al.}, 1994\,; {\it Massonnet et
al.}, 1994; 1996]. The observed postseismic relaxation, with concentration in the
vicinity of the fault, is attributed either to viscoelastic relaxation in the
volume of the crust and upper mantle or to afterslip on continued slip on the fault
planes rupture in the earthquake. There are no particular difficulties to 
account for these
observations within the current understanding of the crust rheology. 

We mention another phenomenon occurring at much smaller scales\,: horizontal fault
slippage has been detected in several faults around the world 
([{\it Bolt}, 1993], pages 92-93), including the San
Andreas fault system, the north Anatolian fault in Turkey and along the Jordan
Valley rift in Israel. The slip is usually highly localized on a width of one
meter or less and has an amplitude of a few millimeter and a duration of a few
minutes to a few days. The standard interpretation is to invoke the weak strength
of the gouge material present within faults in the upper layers of the crust which
can not resist the tectonic applied shear. If the shear is applied more at the
bottom of the crust, as in several model calculations, it is hard to argue that
the upper gouge would deform while the lower stronger layer is stuck. We have
seen that the upper layers respond and follow the after-earthquake continuous
deformation. The observed localized slip at the upper surface
should thus reflect a process occurring in the stronger rock materials,
namely very localized deformations. It is still an open question whether the
faults having measurable creep could carry very large earthquakes (magnitude $5$
earthquakes are been measured on these faults).

\section{The stress paradox \label{friction}}

\subsection{Statement of the paradox}

In the laboratory, frictional sliding is observed to occur at lower stresses than
cracking. This is the usual proposed rational for considering that earthquakes are
better described as slip events than as rupture processes.

There is a large body of litterature
documenting the maximum shear stress to initiate sliding as a function of normal
stress for a variety of rock types. The best linear fit defines a maximum
coefficient of static friction $f_s = 0.85 $ [{\it Byerlee}, 1977], with a range
maybe between $0.6$ to $1$ approximately \footnote{Heuze [1983] reviews a large body
of the litterature until that time on the various properties of granitic rocks.}. The
range has been confirmed by stress inversion of small faults slip data in situ with
result $0.6 \pm 0.4$ [{\it Sibson}, 1994]. The stress inversion methods follow the
assumption that slip along a fault occurs in the direction of the maximum resolved
shear stress. These methods calculate the stress axes which minimize the angular
deviation between the observed slip axis along the fault and the axis of the maximum
resolved shear stress determined from the general stress. 

There is big discrepancy between this value of friction measured for rocks in the
laboratory in situ and the corresponding stress threshold on the one hand and the
stress that is available in nature to trigger an earthquake on the other hand.
Indeed, assuming a Mohr-Coulomb relation, for a normal stress of the order of the
lithostatic pressure, say $300 ~MPa$ at $10 ~km$ depth, the shear stress at
threshold is thus of order $200 ~MPa$ {\footnote{This must be compared 
to the average stress drop for large and great earthquakes which varies between
$1$ and a few $10~MPa$ [{\it Kanamori}, 1994].}}. Using a shear modulus of $10 ~GPa$, this
yields a shear strain at threshold of $\epsilon \approx 3 ~10^{-3}$. This is an
enormous value that would demand a very large localization. Consider a large
earthquake with  rupture slip of $u \approx 5~ m$, then the localization width must
be ${u \over \epsilon} \approx 1.7 ~km$. As already discussed, this is not usually observed.
In addition, the stress of the San
Andreas fault zone deduced from a variety of techniques is found to be low and
directed at a large angle close to perpendicular to the fault [{\it Zoback et al.},
1987]. This is in contradiction with the frictional sliding model. A more detailed
tensorial analysis assuming hydrostatic pore pressures gives the depth-averaged
shear strength of faults in the brittle continental crust under a typical 
continental geotherm and strain rate of about $35~MPa$ ($350~bars$) in normal faulting,
$150~MPa$ in thrust faulting and $60 ~MPa$ in strike-slip faulting [{\it Hickman}, 1991].
Computer simulations modeling fault networks in California with thin-plate finite
elements, when tested against measured slip rates on faults, against principal stress
directions and rates of geodetic baselines, suggest that the time-average friction coefficient
of major faults is only $0.17-0.25$, i.e. only $20\%-30\%$ of the value that is assumed
for the friction of intervening blocks [{\it Bird}, 1989; {\it Bird and Kong}, 1992].

In the previous section, we have discussed the paradox that, in many situations,
strain localization is not observed. Strain localization could be naively 
argued to be necessary to increase the stress level necessary to trigger 
earthquakes. The
problem with this argument is that, if the gouge zone is weak, it will indeed
accumulate and localize the strain but the stress will still be small due to the weak
elastic modulus. On the other hand, if the fault gouge is not so weak, the strain will
not be localized and thus the stress will also be weak due to the small strain, as
the deformation is spread over a large domain.
Let us make this quantitative by using a simple 1D model. Consider two
shear springs associated in series of shear elastic
modulus $k_1$ and $k_2$. One extremity of the springs is fixed and the other one
is displaced by the amount $u$. The force  
\be
F = {k_1 k_2 \over k_1 + k_2}~ u
\ee
is the same in each spring and their respective displacement is 
\be
u_1 = {k_2 \over k_1 + k_2} u~~~~~{\rm and}~~~~~u_2 = {k_1 \over k_1 + k_2} u~,
\ee
 which add up to the total
displacement $u$. Note that if the weak zone (represented by spring $1$) is such
that $k_1 << k_2$, then $u_1 \approx u$, {\it i.e.} essentially all the
deformation is localized within the weak layer. However the force is weaker and
weaker as $k_1$ decreases since $F \approx k_1 u$.  In the other limit where $k_1
\to k_2$ (no weakened zone), strain localization disappears ($u_1 =
u_2 = {u \over 2}$) and the force ${k_2 \over 2} u$ is of the same order. 

Let us translate this spring model by considering that the weak spring has a size
$b$ corresponding to the width of the gouge zone and the strong spring has the
width $L$ of a tectonic plate. 
The previous calculation gives the stress $\sigma_1$ within the weak
zone  \be
\sigma_1 = {k_1 \over 1 + {L \over b} {k_1 \over k_2}}~~ {u \over b}~~~.
\ee
Let us take
${L \over b} \approx {100 km \over 1 km}$. If ${k_1 \over k_2} \approx 1$, we get
$u_1 \approx {u \over 100}$ (no strain localization) and 
$\sigma_1 \approx k_1 {u \over L} \approx 10^{10} {5 \over 10^5} = 0.5~ MPa$.
On the other hand, if the gouge is soft, {\it i.e.} $k_1 \approx {k_2 \over
100}$, then $u_1 \approx {u \over 2}$ (strain localization) but $\sigma_1 = k_1
{u \over b} \approx k_2 {u \over L}$, which gives the same order of magnitude\,:
the enhancement of the strain is exactly compensated by the softening of the
gouge.  We thus see that the largest stress drops that are measured (for
instance $30 MPa$ for Landers earthquake of June, 18, 1992, $M_S \approx 7.3$)
cannot be accounted for by a localization of the strain within a soft gouge. This
simple calculation shows the difficulty in having both strain localization
and significant stress build-up by invoking solely a mechanical deformation
process. We could have guessed this result from the start from stress conservation.

A set of stress-field indicators, including 
borehole breakouts, earthquake focal mechanisms
and hydraulic fractures, suggests that many (but not all) active faults are
sliding in response to very low levels of shear stress [{\it Zoback}, 1992a; 1992b].
For instance, 
the maximum horizontal principal stress, as measured by 
borehole breakouts and hydraulic fractures, is nearly perpendicular to the San
Andreas fault in Central California. Within the standard mechanical model, this implies
that the San Andreas fault is sliding at a very low shear stress. 
It is very important to stress that this conclusion 
is reached by using the usual solid friction law and is thus {\it model-dependent}.
Within alternative scenarios, this conclusion is no more necessary. Also noteworthy is
the paradox of large overthrusts [{\it Brune et al.}, 1993], namely that thrust 
faults \footnote{Thrust faulting refers to a mode of sliding during an earthquake
in which one block moves upward with respect to the other block on the other side
of the fault, so as to accomodate a large scale shortening (or compressive) deformation.}
exhibit an orientation too close to the horizontal to obey the usual 
friction law, thus also requiring an anomalously small resistance to friction.
Such low-angle faulting is observed in many places and is in contradiction with Anderson's
classification of faults [{\it Anderson}, 1951] and with the friction 
theory usually used for the other modes of
faulting. The conclusion on this weakness of faults is contrary to 
traditional views of fault strength based upon laboratory experiments and 
creates a serious problem as one cannot rely directly on the knowledge 
accumulated in the laboratory. 

The situation is made even more confusing when one refers to the occasional observations
of very high stress drops ($30~MPa$ to $200~MPa$) in moderate earthquakes [{\it Kanamori}, 
1994]. This would indicate that the stress drop (and therefore the absolute stress level)
can be very large over a scale of a few kilometers. But then we should see large strain
and anomalous heat flux due to frictional heating (see below).

\subsection{Proposed resolutions}

To account for these puzzles, many suggestions have been made. Let us mention in
particular that Chester [1995]
has proposed a multi-mechanism friction model including cataclitic flow, localized
slip and solution transfer assisted friction in order to describe the mechanical
behavior of the transitional regime at mid-crustal depths. His conclusion is that, on
the basis of existing friction constitutive relations, only very thin faults (about
$1~mm$ thick) can display the rate dependent characteristics (such as stress drop)
necessary for initiation of seismic slip to any significant depth in the crust.
Blanpied et al. [1995] also use the multimechanism friction model for frictional slip
of granite. They stress that extrapolating the laboratory results to conditions not
encompassed by the data set (i.e. to approach the conditions in the crust) is
uncontrolled as many mechanisms are competing in a complex way. Even the
fluid-assisted deformation mechanisms, such as solution precipitation creep, often
invoked to explain the apparent low strength of some major faults actually lead to
velocity strengthening stable slip. An honest statement is probably that we do not
have enough informations, in view of the complexity of all the competing processes,
to determine with some certainty the rheology at seismogenic conditions in the crust.

The solution which is often proposed to resolve this problem is to invoke fluid pore
pressure. The presence of a fluid decreases the normal stress exerted on a fault
and thus decreases the driving shear stress that is
necessary to reach the sliding threshold while keeping the most favorable
slipping direction unchanged. This last condition ensures the compatibility with
the in-situ stress inversion measurements.

The problem however remains to find a mechanism for pressurizing fault fluids
above the hydrostatic
 value  $\rho_{water} g h$ towards the lithostatic value
 $\rho_{rock} g h \approx 3 ~\rho_{water} g h$
in short time scales compatible with earthquake cycles. 
Several scenarios have been proposed in which
low fault strength results from high pore pressure. One model invokes the
expansion of fluid pores from co-seismic shear heating [{\it Lachenbruch}, 1980],
but it is hard to understand how this mechanism would work to trigger an
earthquake. Another mechanism uses a mantle-derived source of fluids to maintain
overpressure within a leaky fault [{\it Rice}, 1992]. To work, the amount of fluid
must be a function of time in a way which is poorly constrained. Another mechanism
is suggested by recent laboratory sliding experiments on granite which show that the
sliding resistance of shear planes can be significantly decreased by pore sealing
and compaction which prevent the communication of fluids between the porous
deforming shear zone and the surrounding material [{\it Blanpied et al.}, 1992].
The problem however, not tested in the laboratory experiment, is to understand
what causes the change in permeability and the sealing of the pores (which are
imposed exogeneously in the experiments). Several works have explored the
mechanisms and characteristic time scales of pore and crack sealing, with the
conclusion that crack sealing occurs more by pressure solution with mass transfer
with time scale of years to centuries rather than by fast self-healing [{\it
Gavrilenko and Gueguen}, 1993; {\it Renard et al.}, 1997; 1998; {\it Gratier et al.}, preprint].
While at first view, this  mechanism seems reasonable, it is at variance with other
well-documented effects of fluid on rocks under stress, namely hydrolytic weakening
and stress-corrosion subcritical crack growth [{\it Scholz}, 1990]. It is thus hard
to justify the postulate [{\it Blanpied et al.}, 1992] that ``a fault remains
sealed indefinitely once pressure seals are in place''. This assumption is also in
disagreement with the relatively high values of crust permeability inferred from in
situ phenomena which indicate that the rocks are in general close to hydrostatic
[{\it Brace}, 1980], implying that water at depth is sufficiently connected to the
free surface.  Similar weakening of the shear strength may
also be attained by the presence of a relatively soft gouge layer surrounding the
harder rock [{\it Lockner and Byerlee}, 1993]. However, in this last case, the low
apparent friction is no more compatible with the orientation of the fault and
in-situ stress measurements. Another problem is, as already mentionned, that
fluid pressure close to lithostatic value implies fluid trapping and absence of
connectivity with larger reservoirs and the upper surface. Thus, the
overpressurization can only work in small local fault regions, which become
similar to open cracks. It is hard to conceive that the
overpressurization will invade the whole length of the fault, in view of the amount
of available fluid. The friction would thus be decreased or even suppressed
only on localized region and it is not obvious that this would decrease
significantly the effective macroscopic friction.

To summarize, it seems that invoking fluid overpressure to weaken the effective
friction coefficient serves a single purpose and creates novel difficulties to
interpret other observations. For instance, how to rationalize within the fluid pore
pressure scenario the observation by Kanamori et al. [1992] and others that the longer
the interval between two earthquakes on the same fault, the larger the stress drop?

\section{The heat flux paradox}

\subsection{Statement of the paradox}

The heat flow paradox [{\it Hickman}, 1991] in a seismically active
region was first proposed by Bullard [1954]\,: to allow for large earthquakes, a
fault should  have a large friction coefficient so that it can store a large amount of
elastic energy and overpass large barriers. However, if the dynamical friction
coefficient is large, large earthquakes should generate a large quantity of heat
due to the rubbing of the two surfaces, which is
not easily dissipated in a relatively insulating earth. Under repetition of
earthquakes, the heat should accumulate and either result in localized melting
(which should inhibit the occurrence of further earthquakes because slow continuous sliding
should then occur) or develop a high
heat flow at the surface. 0bservations over the entire state of California
have shown the absence of anomalous heat flow across the major faults [{\it Henyey
and Wasserburg}, 1971; {\it Lachenbruch and Sass}, 1980; 1988; 1992; 1995;
{\it Sass et al.}, 1992].

Quantitatively, consider a 1906-San Francisco earthquake releasing about
$E \approx 10^{17} ~J$ \footnote{This is probably an upperbound
because it is estimated by assuming that the stress drop is equal to the shear stress
before the earthquake [{\it Knopoff}, 1958], while it is generally thought that the
stress drop is a fraction of the pre-rupture stress.}
over a rupture length of $400 ~km$ and occurring every
$T \approx 100 ~years$.  Assume that a major fraction is dissipated during the
rupture in a localized domain around the fault due to friction, rupture, and
similar processes.  To get an upperbound, we suppose that all the energy is
converted into heat. Neglecting the finite width of the lithosphere, the
one-dimensional heat diffusion problem is easily solved  [{\it Turcotte and
Schubert}, 1982] (page 188). An impulsive heat pulse diffuses with the heat
diffusion coefficient  $\kappa= {k \over \rho c}$, defined
in terms of the coefficient $k \approx 2 ~W m^{-1} K^{-1}$ of thermal
conductivity, the density $\rho$ and the specific heat $c$ of the crust. 
For the lithosphere, $\kappa \approx 1 ~mm^2/s$ typically. 
Over the repeat period of $100 ~years$, an impulsive source of
heat diffuses over the typical scale $\sqrt{\kappa t} \approx 50 ~m$, which is
significantly smaller than the broken zone width (about a few hundred meters to a
few kilometers in a mature fault such as the San Andreas). 
It thus makes sense to use a continuous approximation
(low-pass band filter) and  consider that the amount of heat generated by a series
of repeating earthquakes with period $100 ~years$ is equivalent to a localized
continuous heat source  with output power $q_0$ equal to the average of the
impulsive contributions from earthquakes. This yields 
\be
q_0 \approx {E \over T S}~.
\ee
 Using $E \approx 10^{17} ~J$, 
$T \approx 100 ~years \approx 2.5~~10^9 s$ and $S = 400~km \times 15~km$ (heat is
generated in a narrow zone along the rupture surface of depth approximately given by
the brittle crust depth), we get 
$q_0 \approx 67 ~mW/m^2$. Neglecting dissipation, we can calculate the surface
heat flux $q(x)$ perpendicular to the fault, which is generated by this constant
localized source $q_0$. It is given by $q(x,t) = q_0$ erfc $({x \over
2\sqrt{\kappa t}})$, where erfc is the complementary error function. In the
presence of the upper crust surface, this horizontal heat flux must also
translate into a vertical heat flux of the same magnitude. We thus see that one
should expect an anomalous heat flux of large amplitude localized in the vicinity
of the fault. Our approximations provide an upper bound for this anomalous flux,
equal to $q_0 \approx 70 mW/m^2$. We get the same order-of-magnitude estimate by
following [{\it Lachenbruch and Sass}, 1980; 1988; 1992; {\it Lachenbruch et al.}, 1995] 
and calculating the frictional heating
as the work $v~\sigma$ per unit time and unit volume 
done by the fault moving at its average tectonic velocity $v$ (a few
centimeters per year) under its average stress $\sigma$. This is also the same
as the work $V~\sigma~{\delta t \over T}$ per unit time and unit volume done by the fault
moving during an earthquake at particle velocity $V \approx 1~ms^{-1}$ over a time
$\delta t = {\delta u \over V}$ to produce the slip $\delta u$.
The normalizing
factor ${\delta t \over T}$ accounts for the fact that this heat dissipation only
occurs during the brief instant of time $\delta t$ of the duration of the 
earthquake compared to the large recurrence interval $T$ between earthquakes.

The estimation $q_0 \approx 70 mW/m^2$ is a significant effect, comparable
to the mean surface heat flux in California equal to about $50~mW/m^2$.
Over one hundred heat flow measurements in the San Andreas fault zone give no
evidence for any local frictional heating on the main fault trace at any
latitude over a $1000~km$ length. These measurements provide an upper bound for the frictional
resistance allocated to heat production of about $10~MPa$ [{\it Lachenbruch}, 1980;
{\it Lachenbruch and Sass}, 1980; 1988; 1992; {\it Lachenbruch et al.}, 1995].

\subsection{Proposed resolutions}

Since the previous estimations are upper bounds, we have probably overestimated the
fraction of energy allocated to frictional heating and this is the first possible
explanation. But present measurements are able to detect anamalies of less than 
$10~\%$ of the average flux and this should be enough to access to the frictional heating.

Another standard explanation is that no significant heat is generated because the
dynamical friction coefficient is low on the most rapidly slipping faults in
California. Two classes of models are then usually proposed. In the first class, 
the low dynamical friction is produced during the event itself. Various mechanisms
are invoked, such as crack-opening modes of slip [{\it Brune et al.}, 1993; {\it
Anooshehpoor and Brune}, 1994; {\it Schallamach}, 1971], dynamical collision effects 
[{\it Lomnitz-Adler}, 1991; {\it Mora and Place}, 1994; {\it Pisarenko and Mora},
1994; {\it Schmittbuhl et al.}, 1994], self-organization of gouge particles under
large slip [{\it Lockner and Byerlee}, 1993; {\it Scott et al.}, 1994; {\it Scott},
1996], acoustic liquifaction [{\it Melosh}, 1996] and so on. 

In the second class of models, the fault has a low friction before
the onset of the event. This may be due to the presence of low-strength clay
minerals such as Montmorillonite (the weakest of the clay minerals) [{\it Morrow
et al.}, 1992], to an organized gouge structure similar to space filling bearings
with compatible kinematic rotations [{\it Herrmann et al.}, 1990], to the presence
of phyllosilicates in well-oriented layers [{\it Wintsch et al.}, 1995] or
to the existence of a hierarchical gouge and fault structure
leading to renormalized friction [{\it Schmittbuhl et al.}, 1996].

Much attention
has also been devoted to the role of overpressurized fluid [{\it Lachenbruch}, 1980;
{\it Byerlee}, 1990; {\it Rice}, 1992; {\it Blanpied et al.}, 1992; 1995; {\it Sleep
and Blanpied}, 1992; {\it Moore et al.}, 1996]\,: if water, believed to be
ubiquitous within the crust, is connected to the surface, its pressure is given by
the hydrostatic value $\rho_w g h \approx 100 MPa$ at $10 ~km$ depth ($\rho_w$ is
the density of water); on the other hand, if pockets of water are trapped, the fluid
pressure in these pockets becomes equal to the lithostatic value $\rho g h$, where
$\rho$ is now the density of rocks. This leads to a pressure about three times
larger  $\rho g h \approx 300 MPa$  at $10 ~km$ depth.
In this situation where the fluid pore pressure is elevated to values
close to the normal compressive stress (lithostatic pressure), the fault could be
weakened enough to exhibit a weak stress threshold and friction. Recall that, in
the friction model, a fault is strong if the normal stress is strong to hold the
two surfaces of the fault together. If the interstitial fluid pressure decreases
significantly the normal stress, the fault becomes weak.

Fluid pressure close to lithostatic value implies fluid trapping and absence of
connectivity with larger reservoirs and the upper surface. Thus, the
overpressurization can only work in small local fault regions, which thus become
almost free, similarly to open cracks. It is hard to believe that the
overpressurization will invade the whole length of the fault, in view of the amount
of fluid that would entail. The friction would thus be decreased or even suppressed
only on localized region and not on the whole fault. In addition, laboratory data
[{\it David et al.}, 1994] suggest that a relatively high influx of fluid from
below the seismogenic layer is necessary to maintain pore pressure at close to the
lithostatic value, as postulated in the Rice model [{\it Rice}, 1992]. Scholz
[1992b] has noticed that the fluid pressure scenario has one major problem that
remains to be resolved, namely that a brittle material can never resist a pore
pressure in excess of the least compressive stress $\sigma_3$ without drainage
occurring by hydrofracturing. $\sigma_3$ could then be the fluid threshold of 
a rupture instability which could possibly trigger an earthquake. However, this
mechanism does not lead to a small friction coefficient.
Segall and Rice [1995] have developed a model of unstable
slip of a fluid infiltrated fault using a rate and state dependent friction model
with dilatancy and pore compaction. For earthquake to nucleate, they find that the
fluid pressure must not be large and this places contraints to rationalize the
absence of heat flow anomaly on the San Andreas fault. Furthermore, their numerical
simulations do not exhibit large interseismic increases of the fluid pressure, in
contrast to the often proposed scenario.

There is evidence that the main form of dissipation of the released
tectonic elastic energy is not due to frictional heating. The total elastic energy
released during an earthquake can be estimated using the method of static
elasticity. The idea is to calculate the energy
difference of the crust in two states, before and after the earthquake. The
earthquake is modelled by the introduction of a shear fault with a stress on it
which is reduced with respect to the preexisting uniform shear stress by an amount
equal to the earthquake stress drop. The
total energy released is a quantity which is not measurable directly but
nevertheless is the reservoir of the various possible forms of energy dissipation
that take place during an earthquake. The total dissipated energy comprises seismic
wave radiation energy, which is the usual measured ``earthquake energy'' by seismic
methods, the heat dissipated in friction (not directly accessible), 
the energy consumed in new cracking and
fragmentation in the gouge [{\it King}, 1983;1986] (not directly accessible)
and in the growth of the fault
[{\it Andrews}, 1989] (not directly accessible),  
as well as other relaxation phenomena, including fluid flow
and ductile deformations and possibly polymorphic transformation, chemical and 
electrical processes. Friction, ductile deformations and processes with latent heat
would generate heat, while radiation carries the energy far away and the rupture and
fragmentation consume energy by the creation of new surfaces. The scenario of a low
heat generation thus leads to a large dissipation in the other modes, {\it i.e.} seismic
radiation and fresh rupture. Measurements of the radiated seismic
energy give uncertain values, due to the need to correct for attenuation and
propagation dispersion [{\it Kanamori et al.}, 1993; {\it Choy and Boatwright},
1995].  For Loma Prieta 1989 earthquake, $M_S = 7.1 \pm 0.3$, the 
radiation energy is approximately
$E_S = 10^{15} -  8~~10^{15} J$, which must be compared with the
static elastic calculation using a simple shear model [{\it Knopoff}, 1958] giving
a total released energy   
\be 
E_T \approx  {\pi \over 8} G d^2 L~.
\ee
For a slip $d \approx 1.6~ m$ over a rupture length $L \approx 40 ~km$, we get 
$E_T \approx 1.3 ~10^{15}~J$, which is in the low end of the interval for $E_S$. 
Similar results are found for other large earthquakes. This indicates that
the dissipation by seismic radiation is not negligible and could be a dominant
mode. Kanamori and Anderson [1975] have proposed that the seismic radiation
efficiency is given by $2 {\sigma_f - \sigma_d \over \sigma_f + \sigma_d}$, {\it
i.e.} the ratio of the stress drop over the average stress. In the standard
picture, a large seismic radiation would thus need a small absolute shear stress.
This seems in contradiction with the stress drop measurements in the range $0.1-10
MPa$ which are much smaller than the typical lithostatic stresses in the range of
hundreds of $MPa$. Recent stress measurements by boreholes near active faults suggest
near-complete stress drop [{\it Barton and Zoback}, 1994].
We note that a large radiation efficiency has
important consequences on the form of the organization of earthquakes and their
repetition [{\it Xu and Knopoff}, 1994].

Other suggestions to explain the absence of an anomalous heat flux near active faults
include geometric complexity in the San Andreas fault at depth, hydrothermal circulation
and missing energy sinks (see [{\it Hickman}, 1991] and references therein).

\section{Water and rocks\,: Metamorphism and faulting}

\subsection{The ubiquity and importance of water}

It is more and more recognized that fluids play an essential role in virtually all
crustal processes. The solubility of water is typically $0.15 \%$ by weight in
quartzite. Clay can contain up to $5-10\%$ of water.
Hickman et al. [1995] review the historical development
of the conciousness among researchers of the ubiquitous presence and importance
of fluids within the crust.
Numerous examples exist that demonstrate water as an active agent
of the mechanical, chemical [{\it Wintsch et al.}, 1995] and thermal processes that
control many geologic processes that operate within the crust [{\it National Research
Council}, 1990]. It is now well-established that the presence of water governs
essentially all the physical processes in quartz, from the piezoelectric properties
to dislocation mobility and to recrystallization. Older experimental results published
before this recognition must thus be reassessed with caution. In other words, the
mechanical properties of quartz, and probably of other quartzite minerals, are not to
be explained in terms of intrinsic mechanisms. In the same vein, we will review below
the fact that phase transformations documented in thermodynamic equilibrium may be
drastically modified in the presence of water and deformation, thus enriching
considerably the possible behaviors of rocks.

The bulk of available information on the behavior of fluids comes
from observations of exposed rocks that once resided at deeper crustal levels. These
outcrops give a series of ``snapshots'' (with often multiple exposures) that have the
disavantage of not providing direct samples of fluids and of its role in crustal
development. In any case, present day surface exposures indicate that, at all crustal
levels, fluids have been present in significant volume. Exposed metamorphic 
\footnote{A rock whose original mineralogy, texture or composition has been
changed by the effects of pressure, temperature or loss of chemical components.} rocks
indicate large fluxes of fluids. Fluids have been directly sampled
at about $11 ~km$ by the Soviets at the Kola Peninsula drillhole. All these
data collectively support the existence of fluid circulation and therefore porous
rocks and/or fractures to crustal depths of at least $10-15 km$. Studies indicate
that chemical modification which require significant fluid volume has occurred,
contrary to inferences based solely on mechanical properties. The
advection of chemical components by fluid flow and dispersive fluxes from one
chemical environment to another causes chemical reactions between minerals and
fluids, such as dissolution, precipitation, ion exchange and sorption. The closed
nature of many mineralized fractures illustrates the important interaction between
the chemistry and physical transport properties of the crust. It is recognized
that moving fluids that transport both heat and mass facilitate the precipitation
of minerals, dissolution of the rock matrix, ion exchange, and a host of potential
chemical reactions. The system is often considered dynamic with reactions
occurring in the moving fluid. The fluid can dictate the style of metamorphism,
the extend of chemical equilibrium and much about the texture fabric observed in
rocks. The evidence that large amounts of fluid flow occurred in deeply buried
rocks consists of documented chemical changes in mineral assemblages. Because the
porosity of metamorphic rock is probably less than $1 \%$, the high calculated
fluid necessary to produce observed chemical changes suggests that fluid must have
been replenished thousands of times. If fluid flux through the metamorphic rock is not
continuous, many generations of microcracking will form. Sibson [1982] and Sibson
et al. [1975] document the liquifaction of fault, {\it i.e.} the release of fluids
during and after faulting.

It has been argued that rock permeability and
thus microcracking adjusts itself, so that fluid pressure is always close to rock
pressure irrespective of the extend of hydration/dehydration [{\it Walther}, 1990].
However, the mechanism by which this would occur is not clear. To operate in this
state, feedback processes must exist. In this vain, Nur and Walder [1990] have
proposed a time-dependent process to relate fluid pressure, flow pathways and fluid
volumes. This is in order to reconcile the competing ideas of high crustal
permeabilities and associated hydrostatic fluid pressures with evidence for
hydraulically sealed low-permeability rocks with elevated fluid pressures. They
assume a tendancy of sealing and healing which produce a progressive decrease of
the porosity. If the permeability decays faster than the available volume, fluids
will be trapped and will be raised to the lithospheric pressure. This high pressure can
then break the surrounding rocks in tension, freeing the fluid which depressurizes
down to the hydrostatic pressure. Because the process of porosity reduction continues,
the fluid pressure will again build up towards lithostatic, leading to another cycle of
hydrofracture, fluid release, and sealing, etc. One must take in addition the release
of water from transformed minerals. For instance, Montmorillonite changes to illite
with a release of free water from the clay structure at approximately the same depth
as the first occurrence of the anomalous pore pressure [{\it Burst}, 1969]. This is
the most commonly discussed example of hydration and dehydration of minerals changing
the fluid mass and the pore pressure. It has been proposed that gold-quartz vein
fields in metamorphic terranes provide evidence for the involvement of large volumes
of fluids during faulting and may be the product of seismic processes [{\it Robert et
al.}, 1995].

One can suggest an alternative
explanation for the apparently opposed views of a large permeability on the
one hand and sealed high pressure fluid on the other hand, which is {\it
heterogeneity}. It is clear that the crust is very heterogeneous and the previous
scenario neglects this heterogeneity. In the presence of a large distribution of
pore and fracture lengths and their opening sizes (Walmann et al., [1996] document
a power law relationship between crack opening in tension and crack length), the
permeability at large scale is controlled by the large size fraction of the
population of openings, while trappings and sealing can occur in the dead ends and
small cracks. The two processes can thus coexist without invoking additional
mechanisms.

Direct evidence of the effect of fluid on fault stability has come from
earthquakes induced in intraplate regions either through direct injection
of fluids down boreholes or from the filling of large reservoirs with subsequent
infiltration of water into the underlying rock mass (distinct from the direct
loading effect of the reservoir water mass [{\it Grasso and Sornette}, 1998] and
references therein). Such induced seismicity by reservoir
impoundement and fluid injection could also be a signature of the weakening effect
of fluids on rocks.

The magnitude of fluid pressure transients associated to earthquake faulting have
been estimated from the examination of fractures which exhibit different phases of
induced dilation by escaping fluids [{\it Fleischmann}, 1993].

\subsection{Effect of water on minerals\,: hydrothermal alteration under stress}

Polymineralic crystalline rocks, comprising most of the lithosphere, exhibit
complex strain partitioning that has been little studied but is probably very
important in the chemical-mechanical-electrical processes occurring in the crust.
Kirby [1984] concluded in his digest of the papers of the special issue on
chemical effects of water on deformation and strengths of rocks that it would be
premature to establish with confidence the rheological laws for any crustal rock
type under hydrothermal conditions on the basis the presented result. This
conclusion still holds by large more than ten years later. He also stressed the
fact that the chemical effects of aqueous fluids may be heterogeneously
distributed in the crust. Our understanding of the possible interplays between
earthquakes and rock transformation is still in its infancy.

\subsubsection{Role of water in the presence of cracks\,: stress corrosion}

When the chemical environment of a solid is changed, and some active species adsorbs
on its surface, then because the electronic structure of the near-surface region of
the solid is changed, so are its mechanical properties. This chemomechanical effect
are been studied for a long time for metals [{\it Westwood and Ahearn}, 1984]. The
effect of water on cracks operates similarly.

The presence of water enriches the elastic rebound model of section 2. The
simplest scenario for the preparation of an earthquake which takes into account
the presence of water is the following\,: first, the slow straining of the crust due
to tectonic forces produces many microcracks throughout the rock. Water diffuses
into the cracks and fill them. Water can in fact be also at the origin of the
formation of tension microcracks if sealing occurs in the permeative fluid network
allowing the fluid pressure to increase towards lithostatic values.
During this period, the volume of the region dilates. In fault
zones, these modification and damage brought to the rocks first weaken them by
various processes including stress corrosion. A threshold is eventually reached at
which a major crack can develop along the fault. In this way, the elastic rebound of
the strained material can occur during an earthquake. 

As yet, this sequence of
events has never been directly observed before an earthquake. The scenario is more
based on extrapolations of many laboratory experiments showing the progressive
weakening of rocks with preexisting cracks in the presence of water. Other
experiments have shown the localization properties of rocks in narrow shear bands
under confining pressures comparable to those found in the seismogenic crust. 
These are the elements that are usually advanced to justify the friction model,
which includes the action of water under the form of mechanical
overpressure and of chemical attack, as in corrosion and hydrolytic weakening.

Stress corrosion occurs in the presence of pre-existing cracks
in quartz, quartz rocks, calcite rocks, basaltic rocks, granitic rocks and many
other geological materials [{\it Atkinson et al.}, 1981; {\it Atkinson}, 1984]. Only
limited data exist on crack growth in polycrystalline minerals, but it is clear
that grain boundary chemistry and structure dominate the process [{\it Freiman},
1984]. Several possible mechanisms of subcritical crack growth are relevant
to crustal processes. They include stress corrosion, diffusional mass transport,
microplasticity, dissolution and ion exchange. Below $250-300 ~^{\circ}C$, stress corrosion is
thought to be the main mechanism. Above $300 ~^{\circ}C$, significant crack healing occurs [{\it
Atkinson}, 1984]. There is no data under conditions that simulate realistically those
in the bulk of the earth's crust. In particular, a question can be raised on the
existence of other competing processes involving genuine chemical reactions or
substitutions [{\it Wintsch et al.}, 1995]. Extensive healing occurs at temperatures
as low as $400 ~^{\circ}C$ in times as short as $1-2$ days in presence of fluids [{\it Smith
and Evans}, 1984]. The evidence is that cracks heal by local transport of silica by
diffusion along the crack surface and/or through the pore fluid.

The subcritical crack growth induced by stress corrosion has been
essentially documented for mode I tension stress. Such loading condition can emerge
when a crack is sheared and microcracks in tension then tend to grow at the tip in a
direction parallel to the main principal direction [{\it Scholz}, 1990, p. 27]. As
discussed above, pore sealing can increase the fluid pressure and lead to tension
microcracking. Extrapolation to shear cracks is an open question.
More generally, it is important to note that, while almost all quantitative
rheological data for rocks have been obtained from axi-symmetric compression
experiments, it is widely accepted that most natural rock deformation phenomena are
non-coaxial, such as in fault shear zones. Franssen and Spiers [1990] show from
experiments on polycrystalline salt that there is no justification that flow laws
derived from coaxial experiments can be generalized to non-axi-symmetric
deformations. The anisotropy is in addition enhanced by the development of
anisotropic mineral textures, such as crystallographic preferred orientations
within a deforming material or by microstructural changes such as dynamic
recrystallization or grain boundary alignment.

\subsubsection{Hydrolytic weakening}

In triaxial experiments under dry conditions, it is seen that 
quartz deformation is not observed, in disagreement with the large
deformations of quartzites \footnote{A very hard, nonfoliated, white metamorphic
rock formed from a sandstone rich in quartz sand grains and quartz cement.} seen at
relatively low temperatures. Indeed, quartz appear to have deformed in crustal
rock samples while other adjacent minerals, known to be weaker in the laboratory, have
not.  As discussed by [{\it Goguel}, 1983] and others, what is missing is the
influence of fluids. Hydrolytic weakening (the weakening of quartz by included water)
could well be the resolution of this enigma and there is much continuing research to
quantity this phenomenon. Water alters the yield strength of quartz, reducting it
from $2~GPa$ to $200~MPa$ at $500~^{\circ}C$. It is found that water molecules are the
dominant hydrogen containing species in synthetic quartz and this water is mainly
in tiny atomic inclusions not large enough to form ice when cooled [{\it Aines and
Rossman}, 1984]. It is difficult to extend our imperfect knowledge of hydrolytic
weakening in the laboratory to {\it in situ} conditions [{\it Ranalli}, 1987].

Water may have a chemical effect on minerals directly, without the need for
preexisting cracks. The micro-cracks may even result from the chemical action of
water according to the three mechanisms of 
silicon-oxygen bonds replaced with hydrogen bonds, dissolution of chemical
impurities and ion exchanges.
It is well-documented that the strength of wet rocks is less than the strength of
dry rocks under most crustal conditions. For example, marble masons know that
marble can be deformed continuously using a humid sponge! Hydrolytic weakening denotes
the general softening effect of water on rocks and minerals. Its seems to act above a
critical temperature which is inversely related to water content. The observed
$(p,T)$-dependence of weakening in single crystals is consistent with the
variations of water solubility and diffusivity in quartz. 

Griggs and his team [{\it Griggs and Handin}, 1960] and many other groups have
developed a wealth of laboratory experiments of rock deformation under high confining
pressure and temperature. Modern experiments have shown that, in the absence of
non-hydrostatic stress, it is not possible to cause hydrolytic weakening of dry
quartz [{\it Kronenberg et al.}, 1986]. This is due to the fact that the penetration
of water is extremely small in pure dry quartz and hence the solubility and
diffusivity of water in bulk quartz is very low. Molecular water in
tiny fluid inclusions gives rise to a broad absorption band while hydroxyls with
precise allocation in the crystal lead instead to polarised sharp absorption peaks.
The exact position of these peaks depends on the environment of the vibrating $OH$
and this has led to quantitative estimates of the $H/Si$ ratio in wet quartz and
other minerals [{\it Paterson}, 1982].  According to recent measurements
[{\it Gerretsen et al.}, 1989], the solubility of water in quartz is probably
significantly less than $100~$H per $10^6~$Si and this does not appreciably change
with heating (up to $1100~K$ or pressure build-up up to $1.5~GPa$).

Water diffusivity in quartz appears to be strongly dependent on pressure, stress
and temperature. Also, the solubility of water in quartz increases with
temperature and pressure. It is also probably dependent on the deviatoric stress,
{\it i.e.} on the anisotropic distorsion of the crystal mesh. Note that the effect is
probably weak for quartz, which has a weak anisotropy as measured by the elastic
coefficients or the energies of dislocations as a function of their orientation. For
application to the conditions reigning in the earth crust, the presence of
numerous interfaces and defects is a favorable circumstance for the exchange of
water within minerals. It must also be recognized that water is a very special
fluid [{\it Poole et al.}, 1994; 1993; 1992; {\it Sciortino et al.}, 1990] and can
be structurally, molecularly or gel bound [{\it Bambauer et al.}, 1969]

A ductile deformation of quartz, which occurs due to the activation of
slip systems, is linked to the hydrolytic weakening effect and also probably to
distorsion and structural transformation in the minerals. The transformation
temperature increases with hydrostatic pressure and applied stress but decreases
with water content. The change of slip system by weakening of quartz
crystal by diffusing in water and increasing temperature is coincident with the
hydrolytic weakening temperature [{\it Blacic and Christie}, 1984]. Observations,
based on the orientations of optical deformation lamellae, indicate a change of slip
systems as well as the deformation mechanism at the onset of hydrolytic weakening.
It has been documented that dislocation kink propagate
easily on adjacent $Si-O-Si$ bridges which are hydrolyzed, permitting dislocation
motion by hydrogen bond exchange. The Peierls'stress is also decreased by hydration.

There is not a complete microscopic
understanding of hydrolytic weakening in the bulk. There appear to be four
possibilities for the action of water [{\it Jones et al.}, 1992].
\begin{itemize}
\item {\it Molecular water residing interstitially in $c$ channels}. Small amounts of
water in minerals can be detected by infrared absorption in the wavenumber range
$4000-2500 ~cm^{-1}$, which corresponds to the absorption of $OH$ stretching
vibrations. 
\item {\it the hydrogarnet defect} involving the replacement of a $SiO_2$ unit
with two water molecules (in other words, this amounts to substitute a $Si$ by four
$H$ atoms); Infrared bands seen around $4500~cm^{-1}$ and $5000~cm^{-1}$ in wet quartz
could be due to this hydrogarnet defect. The stable configurations of the hydrogarnet
effect can be calculated by numerical methods [{\it Purton et al.}, 1992].
\item {\it The Frank-Griggs-Blacic defect} which involves
the hydrolysis of the strong $-Si-O-Si-$ bridges by water molecules\,:
\be
(-Si-O-Si-) ~~~~ + ~~H_2O ~~~ \to ~~~~~ (-Si-OH~\cdot~HO-Si-)
\ee
The detailed impact of this hydrolysis on the glide motion of dislocation by motion of
kinks or the climb and recovery rather than glide is not ascertained [{\it Paterson},
1990]. 
\item {\it Molecular water} found in dislocation cores or at grain boundaries whose
concentration is defect dependent. Dislocation mobility in silicates is
known to be strongly influenced by the presence of structurally bound water. The
concentration of structurally bound water in silicates is dependent upon the
chemical and physical environment.

\end{itemize} 

Discussions of the supposed mechanism of hydrolytic weakening is
often by analogy with dislocation processes in metals, which neglect the charged
nature of defects in minerals which can have significantly different effects on
plastic deformation [{\it Hobbs}, 1981]. Electronic-chemical aspect of deformations in
minerals has received less attention ([{\it Blacic and Christie}, 1984] and references
therein). The electronic defect center theory of Hirsch [1981] and Hobbs [1981] as
well as the Frank-Griggs-Blacic model of a lattice resistance mechanism by dislocation
kink nucleation seem both to have explanatory powers but none adequately explain all
the existing data. The mechanism(s) of hydrolytic weakening is (are) still uncertain
and controversial.

In addition, one must consider the presence of species which can undergo ion exchange
with species in the solid phase. If there is an important mismatch in the atomic size
of the species, lattice strain can result from ion exchange. An example is the
exchange of $H^+$ by $Na^+$ in silicate glasses [{\it Atkinson}, 1981]. 
The majority of silicates, especially the common rock-forming minerals, are
wide-band gap materials, with an optical band gap in excess of about $7 ~eV$
[{\it Nitsan and Shankland}, 1976; {\it Hobbs}, 1984]. Since optical bands
corresponds to relative motions of atoms within the crystal, this high gap
reflects the rigidity of the atomic structure of silicates. Indeed, their
crystal structures are dominated by the $(SiO_4)^{4-}$ tetrahedron. Inclusions of
charged atoms ($Al$, $H$, $Na$, $Fe$ etc) disturb silicon sites and sometimes
decrease the band gap as well as create impurities levels within the gap. Atomic
inclusions act as acceptors or donors of electrons and thus play an important role
in the electric properties as well as structural properties of the silicates 
[{\it Hobbs}, 1984]. For instance, the incorporation of hydrogen, as a donor of
electron, is quite common. The influence of an impurity depends on the mode of
incorporation, as an interstitial or substitutionnal atom or group of atoms in
the crystal structure of the material. The importance of the inclusions stems
from the fact that, in quartz for instance, electrical conductivity is normally
ionic, the charge carriers being interstitial sodium, lithium, etc. The main
hydrogen defect introduced in silicates seems to be the ``hydrogarnet'' defect,
where a $(Si0_4)^{4-}$ group is replaced by a $(H_30_4)^{5-}$ group [{\it Hobbs},
1984]. Hydrogen in minerals most frequently occurs bonded to oxygen. The resulting
$OH$ group is highly polar. We can thus expect a strong coupling between electric
and mechanical structures and properties.

In a series of works [{\it Jones et al.}, 1992; {\it Heggie}, 1992; {\it Heggie et
al.}, 1992], density functional and quantum mechanical calculations of the structure of
water in quartz, its diffusion and the role of dislocations have shown that there are
important changes to the quartz crystalline network caused by the nearby water
molecules. Calculations of defects in quartz are thus likely to be incorrect unless
they take into account the perturbation of the network. The effect is less pronounced
when the water molecule is located in a dilated dislocation core and this leads to a
much reduced activation energy for dissolution and diffusion. Examination of
dislocations in the basal plane of $\alpha$-quartz has shown that the activation
energy for diffusion of water molecules can drop below $1~eV$ due to the interaction
with the dislocation core. The structure of the dislocation cores is itself
influenced by the presence of water molecules. Similarly, reduction in kink-pair
formation energy from $5~eV$ for dry quartz to $1~eV$ or less for wet quartz is
possible, indicating that wet quartz would deform relatively easily, provided that
water could diffuse to dislocations. It has been shown that water can not only diffuse
easily along dislocations in quartz, but also be `pumped' along them [{\it Heggie},
1992]. Numerical simulations indicate that water molecules can dissociate in
dislocation cores and the protons and hydroxyl groups become strongly bound to kinks.
A shear stress can do work on kinks to move them along a dislocation, dragging with
them water-bearing species and sweeping other undissociated water molecules before
them. This process underlines the importance of microcracking as stress enhancement
at crack tips creates and moves wet dislocations which are fed by a reservoir of
water in the crack. The dislocations themselves facilitate the spread of water into
the crystal and in so doing catalyse their own motion.
These observations underlie the interplay
between plastic deformations (involving dislocations) and the presence of water.

In sum, hydrolytic weakening occurs only when defects (such as dislocations) 
are present in rocks and in the presence of non-hydrostatic stresses. Hydrolytic
weakening has mostly been studied for quartz but occurs also in many other
minerals constituting rocks. It belongs to the
general domain of ``chemomechanical'' effects discovered by Rebinder in 1928 (see
[{\it Westwood et al.}, 1981] and references therein) where several basic mechanisms
have been found to operate, none of them sufficiently well-understood.

\subsection{Role of water in rock metamorphism and relation to earthquakes}

There is overwhelming evidence from
geological observation of rock that mineral metamorphism is associated to faulting. 
Faults thus provide an obvious relationship between earthquakes and metamorphism. 
But is the relationship only structural, with earthquakes and metamorphism
occurring on the same geometrical structure at separate times with no real
interaction? Rock metamorphism is the domain of study of geology and petrology. It
involves long time scales in comparison to the relatively short recurrence times
between large earthquakes. 

We summarize a portion of the litterature which presents evidence of the complex
chemical and transformation reactions occurring within fault zones in the
different (relatively rare) parts of the world where evidence has been brought to
the observable earth surface. We note that the available evidence consists in the
observation of shape and chemical patterns, which are the signature of maybe a
succession of processes. The inverse problem of the inference of the causative
factors is obviously quite ill-defined  and unstable. It is thus necessary to
constraint the inversion by a priori information. The interpretations are thus
given {\it within} a priori given physical models, such as friction or rupture. It
is important to keep in mind this procedure for a possible reassessment of the data
in the light of different models and new information. The informations extracted
from the data are thus heavily model dependent.

Chemical effects on deformation are generally not understood so well because of
their manifold and complex nature. These coupled effects are in some instances
cooperative and in others competitive. As as been emphasized by many authors [{\it
Carter et al.}, 1990], metamorphic reaction kinetics and thermally activated
deformation rate processes are tightly interwoven. There is increasing evidence of
strong coupling between deformation and metamorphism [{\it Carter et al.}, 1990].

The metamorphism is known to start at a  minimum temperature of $350 ~^{\circ}C$ but
this depends considerably on the chemical/water conditions.
Evans and Chester [1995] have documented the fact that fluid-driven
reactions, neomineralization and veining occurred during faulting in a small
region near the fault zone less than a meter wide. This extremely narrow fault
core contains ultracataclasite and foliated cataclasite ({\it i.e.} highly
fragmented rocks). Geochemical combined with microstructural and mineralogic data
show that the degree and style of fluid-rock interaction in the fault  core vary
along the strike of the San Gabriel fault.

Hydrothermal mineral reactions can occur at temperatures well below those
necessary for the plastic flow of quartz, and can dramatically lower the ductile
strength of the granite (through presumably basal plane dislocation glide in the
mica) [{\it Janecke and Evans}, 1988]. Stress-enhanced hydrothermal mineral
reactions are also recognized to be important in weakening crustal rocks, even
when both the reactant and product phase are strong [{\it Rubie}, 1983; {\it
Pinkston et al.}, 1987]. Wintsch et al. [1995] have speculated that a hydrated
phyllosilicate-rich fault rock slipping predominantly by dislocation glide may
develop a low and pressure-independent shear strength approaching that of mica
single crystals, without fluid overpressuring, if the phyllosilicate grains are
preferentially oriented and highly contiguous. This reaction softening is promoted
by the presence of $Mg$-rich wall rocks.

Vein textures and rock chemistry indicate that the veins episodically dilated and
collapsed during the history of fluid influx and outflow, suggestive of an
interconnected fracture network that grew either quasi-statically or suddenly
during earthquakes. It must be stressed that stress heterogeneity induced by fault
rupture and deformation can lead to considerable uncertainties in inferring past
fluid pressures from observations of vein geometry in outcrop. No reliable method
currently exists to distinguish the cataclastic products of seismic versus
aseismic slip in fault zones.

Much less data exist on rock-forming silicates (other than quartz) on their 
strength and ductile deformation as a function of pressure, stress, deformation
rate, temperature and water content. This is why, as we have already pointed out,
 most of the evidence on
the effect of water has been documented for quartz. Natural
quartz crystals are very strong\,; by contrast, there is abundant microscopic
evidence in quartz-bearing rocks that quartz has flowed extensively in nature and
was weaker than other minerals, such as the feldspars, under most conditions in the
earth's crust [{\it Blacic and Christie}, 1984]. As explained above, this paradoxical
contrast between high strength of quartz in the laboratory and its apparent weakness
in nature has been qualitatively explained in terms of dissolved water content and the
effect of low strain rates on the flow behavior [{\it Griggs and Blacic}, 1965]. There
is a strong contrast between very strong, dry natural quartz crystals and
their hydrolytic weakening forms.

There are many indications that crystal plasticity and super-plasticity occurs at
pressure and temperature conditions found in the brittle part of the crust. The
plasticity is associated with metamorphism. Fine-grained albite
microstructure formed in feldspar prophyroclasts have thus been documented in
granite and quartz-felspar mylonite \footnote{A very fine grained metamorphic
rock commonly found in major thrust faults and produced by shearing and rolling
during fault movement.}. In these systems, millimetric layer partitioning and grain
sliding in a very-fined-grained mixture have been found [{\it Behrmann and
Mainprice}, 1987]. Microstructures in cataclasites \footnote{High pressure, low
temperature structures occuring primarily by the crushing and shearing of rock
during tectonic movements and resulting in the formation of powdered rock.}, for
instance in basement rocks of the Scandinavian Caledonides, exhibit many forms\,:
microcracks filled with fine-grained mineral fragments, quartz crystals with
euhedral terminations, undulose extinction, deformation lamellae and grain boundary
recrystallization, narrow banding parallel to rhomb and prism planes, vein-like
structures. The interpretation is usually that formation of cataclasites is a
cyclic operation of brittle failure, crystal growth followed by plastic deformation
[{\it Stel}, 1981].  Deformed calcite relics in deformed marbles from the Mt. Lofty
Ranges, South Australia,  show abundant evidence of strain-induced migration of
twins and grain boundaries, together with static and dynamical recrystallization
[{\it Vernon}, 1981]. Fluid inclusions rapidly change shape and re-equilibrate upon
change of temperature and pressure [{\it Pecher}, 1981]. Deformation tests on
aplite at constant strain rate ($10^{-5} ~s^{-1}$), at $210-250 ~MPa$ confining
pressure and at $200 ~^{\circ}C$ and above and at $560 ~^{\circ}C$ and above with varying water
contents show a first plastification regime [{\it Paquet et al.}, 1981]. In several
tectonic settings, a steady-state dynamical recrystallization of quartz grains
has been shown to be essentially independent of post-creep processes. In addition,
the stresses were estimated $1.5-2$ times higher in these zones [{\it Etheridge and
Wilkie}, 1981]. The discrepancy with the stress and friction values found in
experiments has led to the suggestion that laboratory experiments have not been
able to reproduce the conditions within seismogenic fault zones. 

Investigations of
the mineral chemistry of shear zones in amphibolite facies metagabbro has shown
that the composition of the amphibole and plagioclase varies with deformation. The
amphibolite varies progressively from an initial magnesio-horn-blende to a ferroan
pargasitic hornblende in the shear zone, with alkalties, titane, ferrous and
magnesium ions increasing in concentration. Such changes are more usually
associated with increasing temperatures, showing the role of deformation to produce
complex metamorphic and chemical changes [{\it Brodie}, 1981].  As we have described, the
``water-weakening'' effect of quartz has been suggested to be due to the effect of
$OH$ ions which act as electron acceptors and so greatly enhances the diffusivity
of oxygen and of charged jogs on dislocations. Thus, chemical components such as
oxygen, water, aluminium and other ions can control the concentrations and mobilities
and point defects in minerals. For applications to earthquakes, it is noteworthy that
almost all experiments on silicates, sulphides and carbonates have been carried out
under unspecified chemical conditions [{\it Hobbs}, 1981]. It is remarkable that
low-grade metamorphic slates and fold belts show crenulation cleavage which
progressively obliterated with further increase of metamorphism and cleavage with
syntectonic phyllosilicate recrystallization. The fine fabric results from progressive
solution, crystallization and recrystallization superimposed with bending, kinking,
and internal rotation achieved by slip parallel to the basal planes of
phyllosilicates [{\it Weber}, 1981]. Interactions between deformation and
metamorphism in slates leading to cleavage involves mechanical rotations, solution
and crystallization, recrystallization and metamorphic reactions [{\it Knipe},
1981]. Variations of grain chemistry and orientation have been suggested to control
the different deformation behaviors of different grains within the strain
accomodation zones and dictate whether grains bend, kink, dissolve, recrystallize,
react and grow. 

As much as ten potential chemical reactions are been found to be
involved in cleavage, ranging from ionic exchange to crystallization from solution
and solid state reordering. These processes are influenced by deformation with
respect to their location and rate. Large normal fault zones are characterized by
intense fracturing and hydrothermal alteration and the displacement is localized in
a thin zone of cataclasite, breccia and phyllonite surrounding corrugated and
striated fault surface [{\it Bruhn et al.}, 1994; {\it Wintsch et al.}, 1995].
Hydrothermal alteration of quartzo-feldspathic rock at temperature equal or above $300
~^{\circ}C$ creates mica, chlorite, epidote and alters the quartz content. The alteration
processes are of course dependent on the minerals. Intragranular extension
microcracks, transgranular cataclasite extension fractures, through-going
cataclasite filled shear microfaults are observed to result from diffuse mass
transfer and intracrystalline (low temperature plasticity). Young faults consists
of blocks of relatively intact rock separated by narrow zones of intense
deformation where fracture processes dominate [{\it Knipe and Lloyd}, 1994]. A
clear correlation has been observed between cataclastic deformation and mineral and
elemental distribution which has been interpreted to result from a
deformation-induced dolomite to calcite transformation in the $1-2 ~m$ wide shear
zone of intense deformation. The transformation has resulted in removal of magnesium
from the shear zone, selective deposition of calcite as an intergranular cement in
cataclasite/microbreccia units and a relative increase in the concentration of
detrital quartz and feldspars [{\it Hadizadeh}, 1994].

Microstructures and particle size distributions in the damage zone of the San
Gabriel fault imply that deformation was almost entirely cataclastic and can be
modeled as constrained comminution. Cataclastic and fluid-assisted processes are
significant in the core of the faults, as shown by pervasive syntectonic
alteration of the host rock minerals to zeolites and clays and by folds, sheared
and attenuated cross-cutting veins of laumontite, albite, quartz and calcite [{\it
Chester et al.}, 1993]. The structure of the ultracataclasite layer reflects
extreme slip localization resulting from a positive feedback between comminution
and transformation weakening. Optical and scanning electron microscopy of textures
in rocks show that syntectonic alteration of feldspars, the presence of iron
oxides in the faults,  and late-stage quartz veins attest to the flow of water
during deformation and to the syntectonic development of quartz [{\it Evans},
1990]. Fractures in feldspar associated to cataclastic flow lead to the development
of clay-rich cataclasites at low temperature. In feldspars, the silica $SiO_4$
tetrahedra form a three-dimensional framework in which cation sites accommodate
calcium, sodium and potassium ions. In contrast, clays are mainly composed of micas
in which the $SiO_4$ tetrahedra  are linked with cations to form two-dimensional
planar sheets. The transformation from feldspar to mica is an example of the
crystalline transformation occurring in deformation. These transformations occur
preferentially along feldspar grain boundaries and intragranular fractures. Evans
[1990] documents the fact that hydrogen (which has a much larger mobility than
oxygen in silicates) is added to the feldspar by the penetration of water in the
rocks along intergranular fractures and the development of mica and quartz result
in the release of $K^+$ and $Na^+$ ions. 

The potential role of crystal chemistry in determining mechanical behavior in
multicomponent mineral systems has been highlighted by the recent proposed
resolution of a paradox on conflicting data on the relative mechanical strength of
muscovite and biotite [{\it Dahl and Dorais}, 1996]. The basic suggestion is that
interlayer-site energetics of micas exert the dominant kinetic control on
various processes, including volume diffusion and basal slip. The
crystal-chemical resolution of the paradox is based on the fact that 
the different biotite samples which were compared had in fact differing fluorine
$F(OH)_{-1}$ substitutions. The fluorine
substitution and concomitant removal of hydroxyl $H^+$ increases the mechanical
strength of biotite by elimination of $K^+-H^+$ repulsion. This then 
strengthens the interlayer bonds, and by a stronger binding of $K^+$ in the
interlayer cavity, opposes more strongly basal slip.  This is a simple
example where substitution is closely linked with strength and the style of
deformation. Testing of these type of crystal-chemical effects on rheology and
deformation requires mineral specimens of well-controlled composition.

\section{Structure and polymorphism of crustal minerals}

It is also useful to give some informations on the richness of polymorphic phase 
transformations in rocks as they may have some effect on faulting and earthquakes.

X-ray crystallography, spectroscopy, electron-spin resonance and proton-spin
resonance methods have been used to study chemical reactions between two solid
phases [{\it Lonsdale}, 1969]. Such a reaction is in general controlled by local
diffusion of atoms and is most readily occurring in powdered or microcrystalline
specimens and in defect structures having vacant sites, or at the interfaces. A
chemical reaction within a single crystal is characterized by the nature of the
change in the substance itself, with a minimum of atomic or molecular movement, or
by decomposition, isomerization, dimerization or polymerization generally. The
general form of the crystal often remains unchanged so that the final substance is
pseudomorph on the original. The basic constitutive
$SiO_4^{4-}$ tetrahedra unit of many mineral rocks is quite favorable in
this respect to these types of chemical reactions. Indeed, silicate minerals
form complicated structures in which $SiO_4^{4-}$ tetrahedra are linked through
a common oxygen. In this way, it is possible to form polymer chains (as for
instance in the fibrous asbestos minerals); in the micas, two-dimensional
crosslinking of chains form anionic silicate sheets which are held together
through bonds to cations. In the aluminosilicate minerals, such as the feldspars,
the $SiO_4^{4-}$ tetrahedra are linked in a three-dimensional network, where some
of the tetrahedra are replaced by $AlO_4$ tetrahedra.

Silicates are full of structural solid-solid transitions [{\it
Salje}, 1990; 1992]. Among them, displacive transitions are diffusionless
transformation resulting in a local structural modification of the crystalline
structure. The locality of the structural reorganization of the atoms is the reason
for the diffusionless nature of the transitions. More generally, martensite
transformations denotes any diffusionless structural transformation resulting in a
long-lived metastable state with a high degree of short-range order. The existence of
these structural phase transitions is a signature of the many different conformations
that the $Si02$ structural element can take in response to
disturbances created by deformations and inclusions.

Silicate framework structures
consist of rather stiff or `rigid' units, such as $SiO_4^{4-}$ tetrahedra and
$AlO_6$ octahedra, rather floppily jointed at the corners of shared oxygen atoms
[{\it Heine et al.}, 1992]. One can visualize this stiffness constrast by looking
at a two-dimensional perovskite structure\,: in this structure, the
squares in successive layers can rotate alternatively by angles $\pm \theta$ giving
a rigid unit phonon mode in which the structural units rotate and translate without
any distorsion. In quartz and silicates, the tetrahedra are the rigid units which
rotate and translate. In quartz, the tetrahedra rotate by $17$ degrees with a
relative variation of volume ${\Delta V \over V} = 0.1$. Rotations of the units 
reduce the volume of the system, which in turn lowers the Van der Waals
energy of the system and results in a phase transition to a structure with rotated
units. The Van der Waals attraction has an effect similar to pressure,
favoring the decrease of the total volume to bring the atoms closer together. This
effect together with the rigid unit phonon modes explains why silicate framework
materials are riddled with structural phase transitions. The issue of
microdeformation of quartz and other silicates is a complex one. It will remain
controversial unless deformation experiments simulate a very much wider range
of environmental conditions.

The simplest example is quartz, which exhibits several polymorphic phases
that transform into one another by only local atomic motion. Quartz $\alpha$
(rhomboedric system with no center of symmetry and with density $2.65$) is known to
transform into quartz $\beta$ (hexagonal system with a center of symmetry) above
$573~^{\circ}C$ at atmospheric pressure with a transition temperature increasing with
pressure.  The atomic structures of both crystals allow to determine
that quartz $\beta$ is simply obtained from quartz $\alpha$ by a local
rotation of the $SiO_4$ tetrahedra.  The internal tetrahedral interactions play a
dominant role in the transition, as compared to inter-tetrahedral interactions which correspond
to very weak interactions for the rotations [{\it Smirnov and Mirgorodsky}, 1997].
Quartz $\alpha$ also transforms into coesite \footnote{
The structure of coesite (monoclinic system with density
$2.93$) is composed of SiO4 tetrahedrons that are linked 
into four membered rings. The rings are then linked together
into a chain-like structure. This structure is much more compact than the
 other members of the quartz group, except stishovite, and is
reflected in the higher density and index of refraction.}
above $2000 MPa$ at room temperature with a transition pressure increasing
with temperature. The inversion from quartz $\alpha$ to coesite is
not displacive as bonds must be broken in contradistinction with the
transformation to quartz $\beta$. The transformation thus involves atomic
diffusion [{\it Nicolas and Poirier}, 1976]. The arrangement of tetrahedra in
coesite is  very similar to that found in feldspars.

These structural transformations can occur either slowly or
quite fast, depending on the degree of strain concentration and alteration.
Displacive transformations (such as quartz $\alpha$ to quartz $\beta$) can occur very
fast due to the small motions of atoms involved in the
polymorphic transition. The transformation front can propagate very fast, with a
velocity comparable to the velocity of sound, 
as a kind of ``domino wave'' from crystalline cell to its
adjacent cells [{\it Rao and Sengupta}, 1997]. In contrast, sliding or rupture 
involve  deformation instabilities that need to activate large domains within the
minerals and the nucleation of their dynamics is slower [{\it Dieterich}, 1992; 1994; 
{\it Dieterich and Kilgore}, 1994; 1996] as it is
controlled by the mobility of dislocation and other defects.
The kinetics of the transitions but also the stability of the different
phases are also influenced by the presence of water.

\section{The mechanical rupture\,: possible role of high frequency acoustic vibrations}

We now examine the co-seismic rupture. It is expected that 
intense fragmentation and comminution must occur during the sliding,
with the generation of intense shaking and high frequency sound waves. 
This may lead to a kind of partial acoustic liquifaction of the granular gouge, 
similarly to what Melosh [1996] proposed. Indeed, the gouge zone at $10~km$ depth 
can be thought of as a granular medium with 
a large polydispersity in grain sizes, highly compacted by the overburden
lithostatic pressure.

Liquifaction usually refers to the experimental observation
that granular material in the presence of an interstitial fluid can liquify when
shaken sufficiently strongly [{\it Russo et al.}, 1995; {\it Chirone et al.}, 1995].
The liquifaction is due to the fact that granular media are first compressive for
small deformation leading to an increase of the interstitial fluid pressure. This
increase in turn decreases the friction between the grains that can eventually become
free to shear. For example, liquifaction of sediments by resonant amplified
seismic waves have been proposed to be in part responsible for the damage and collapse
of certain buildings during the Michoacan earthquake, 1985 [{\it Lomnitz}, 1987; {\it
Campillo et al.}, 1989]. A similar liquefaction is believed to be responsible for the
the damage in the Marina district of San Francisco during the Loma Prieta earthquake
of 1989 [{\it Bardet}, 1990]. 

Melosh [1996] has suggested that acoustic
liquifaction occurs {\it without interstitial fluids}
 when the local acoustic pressure, due to the intense multiply
scattering high-frequency acoustic waves, can become of the order of the overburden
pressure $\rho g h$, thus letting free a loosely cohesive granular material. 

For an earthquake comparable to the 1906 San Francisco earthquake,
the total released energy is of order $10^{16-17}~ J$. Let us compare this value
with the energy in the acoustic field. The typical acoustic material velocity $V$ (which is
different from the shear velocity) is such that $p =\rho c V = \rho g h$. The first
relation $p =\rho c V$ relates the acoustic pressure to the acoustic particle velocity
($\rho$ is the material density and $c$ the compressional acoustic wave velocity). The
second equality is the condition that the acoustic pressure be equal to the overburden
pressure at depth $h$. This equation states the condition that the acoustic
pressure can liquidify only if it balances the compression. Taking $p \approx
200~MPa$, a density of $3$ times that of water
 and a velocity of order $4000~m/s$, we get the very large
value for the acoustic particle velocity $V \approx 120~m/s$. 

The total acoustic
energy is $p^2/2\rho c^2$ times the volume of the earthquake. If we take a
volume of $400~ km \times~ 10~km$ (active zone depth) 
$\times \sim 1~m$ active fault width, this gives an
acoustic energy of $4~10^{15}~ J$, about one order of magnitudes smaller than the
total energy released by the earthquake. Thus, this mechanism is compatible with the
energy balance since only a (rather significant) 
fraction of the total released energy needs to be
in the form of the high frequency acoustic waves. 

There is however a problem with the Melosh's mechanism because it
predicts a slip velocity during an earthquake more than two orders of magnitudes
smaller than the typical meters per second for observed earthquakes. 
The argument goes as follows.
On one hand, the intrinsic quality factor $Q$ of the attenuation of 
the acoustic waves must be high, in the range of $1000$ or more for the
creation of the acoustic waves to be self-sustained during the earthquake slip
motion (as discussed by Melosh and as seen from eq.(6) of [{\it Melosh}, 1996]. 
Recall that the intrinsic $Q$
measures the dissipation to heat and {\it not} due to scattering, because scattering
is not a loss of energy in the diffusive multiple-scattering transport of the 
high-frequency acoustic energy (eq.(2) of [{\it Melosh}, 1996] that creates the
shaking at the origin of the acoustic fluidization mechanism. 
On the other hand, Melosh estimates
that the attenuation length $l_* = \sqrt{\xi \lambda Q / 4c}$ must be of the
order of the typical wavelength $\lambda$ for the slip velocity 
${\dot u} \approx 1.4~(\tau / \rho c) l_*/\lambda$ to be of the
order of a meter per second. $\tau$ is the
stress drop, $\rho$ is the density,
$\xi \approx c l_e/3$ is the diffusion coefficient [{\it Ishimaru}, 1978]
of the acoustic wave diffusion process in the multiple-scattering regime, 
$l_e$ is the so-called elastic mean free path [{\it Ishimaru}, 1978], 
i.e. scattering length, and $c$
is the transport velocity, of the order of the shear wave velocity 
[{\it Van Albada et al.}, 1991; {\it Van Tiggelen and Lagendijk}, 1993; 
{\it Turner and Weaver}, 1996]. By definition
of the $Q$ factor [{\it Knopoff}, 1964], $Q \equiv 2\pi~E/\Delta E$, where $\Delta E$ is
the energy loss over a wavelength or equivalently a cycle, we get
$Q = 2\pi~ l_*/\lambda$. We thus see that $Q \simeq 1000$ and $l_* \approx \lambda$
are incompatible. We stress again that $l_*$ is distinct from the scattering length $l_e$ 
(which can be itself comparable to $\lambda$ for strong scatterers)
because $l_*$ measures the true attenuation of the high frequency acoustic field and
because scattering which is the mechanism by which the acoustic field is shaking the
gouge is not a dissipation. For the crust, 
we expect indeed $Q \simeq 1000$ [{\it Melosh}, 1996; {\it Knopoff}, 1964] and thus 
$l_* \simeq 160 ~\lambda$, leading to a slip velocity ${\dot u} \simeq 7.5~mm/s$,
using the numerical example for eq.(9) of [{\it Melosh}, 1996].

We may try to save this mechanism by invoking that
the acoustic pressure does not need
to reach the overburden pressure but only a small
fraction $\eta$ of it, in order to liquidify the fault. 
 Indeed, it is well-established experimentally 
[{\it Biarez and Hicher}, 1994] that the elastic modulii of granular media under large cyclic
deformations are much lower than their static
values. This effect occurs only for sufficiently large amplitudes of the cyclic
deformation, typically for strains $\epsilon_a$ above $10^{-4}$. 
At $\epsilon_a = 10^{-3}$, the elastic modulii
are halved and at $\epsilon_a = 10^{-2}$, the elastic modulii
are more than five times smaller than their static values. As a consequence, the
strength of the granular medium is decreased in proportion. Extrapolating
these properties to the crust, we need 
to estimate the strain created by the acoustic field. The acoustic
pressure is related to the acoustic particle velocity $v$ by $p = \rho c v$.
Assuming $p = \eta \rho g h$, this yields $v = \eta g h/c \approx 12~m/s$ for
$p \approx 200~MPa$, a density  $\rho = 3~10^3~kg/m^3$, a velocity 
$c=4000~m/s$ and $\eta = 0.1$. At a frequency $f$, 
this corresponds to an acoustic wave displacement $u_a = 
v/2 \pi f \approx 2~10^{-3}~m$ at $f \sim 10^3~Hz$.
 The corresponding strain $u_a/w$ is $\sim 2~10^{-3}$
for a gouge width $w$ of the order of one meter [{\it Melosh}, 1996] over which the
intense shaking occurs. These estimations suggest that Melosh's criterion
that the acoustic stress fluctuations must approach the overburden stress on the fault
for acoustic fluidization to occur
is too drastic and smaller shaking can reduce significantly the fault friction.

Persuing this reasoning, we see that the fundamental equation (6) in [{\it Melosh}, 1996]
is changed into
\be
{d^2 \Psi \over d\zeta^2} = \Psi - r\Sigma^2 \biggl[ {1 - erf({\eta \over 2\sqrt{\Psi}})
\over 1 + erf({\eta \over 2\sqrt{\Psi}})} \biggl]~,
\label{aaqswq}
\ee
where $\eta = 1$ recovers the case treated by Melosh.
$\Psi$ is the normalized acoustic energy, $\zeta = z/l_*$, $z$ is the coordinate
perpendicular to the fault, the regeneration parameter is $r = eQ/2$ where $e$ is the
acoustic energy conversion efficiency, $\Sigma = \tau/ \rho g h$ and $erf(x)$ is the
error function. This equation expresses the fact that the high frequency
acoustic energy density results from the motion of the fault that creates the shaking
(last term in the r.h.s.) and is dissipated by the intrinsic absorption (first term
in the r.h.s.).

We see that a factor $\eta<1$ implies a more effective
generation of acoustic waves because the second source term of the r.h.s. is larger. But,
since the bracket term saturates to one for large energies and/or small $\eta$,
this does not lead to a significantly 
larger slip velocity than found above. This remains a 
problem of the theory. This problem might be alleviated by
treating $e$ self-consistently 
as a decreasing function of the friction coefficient, and thus of the acoustic energy density.
The problem then becomes even more non-linear because it reflects 
in addition the vibration-weakening strength of the granular gouge. Further 
improvement should also take into account that the stochastic acoustic field is known to be
distributed according to
a Rayleigh distribution [{\it Ishimaru}, 1978] instead of a Gaussian distribution, 
which modifies the functional form of the term in bracket in eq.(\ref{aaqswq}) and
thus all numerical estimations.

\section{Seismic radiation \label{errrrrzz}} 

\subsection{The different regimes}

There can be several contributions to the measured seismic radiation. 
If the earthquake is accompanied by a sudden change of shear modulus localized in the
fault core, Knopoff and Randall [1970] have shown that this transformation
 in the presence of a pre-existing shear strain leads to a double-couple source with the
correct and usual characteristic radiation pattern found in earthquakes at low frequencies.
In fact, it is impossible to discriminate between this source and that of the displacement
dislocation from their first motions. 
The size of the corresponding
double-couple moment is [{\it Knopoff and Randall}, 1970]
\be
M = 2 ~\delta \mu ~\delta \epsilon~ V~,
\ee
where $\delta \mu$ is the change in shear modulus, $\delta \epsilon$ is the change
in strain inside the volume $V$ in which transformation has occurred. For the sake
of illustration, let us take the
change in modulus corresponding to a transformation of quartz to another phase such as
coesite. This is only offered as an illustration of the many possible transformations
that can occur. The bulk modulus of coesite is $9.4~10^{10}~Pa$, compared to that of quartz
which is $3.7~10^{10}~Pa$ [{\it Liu}, 1993] \footnote{The measurement of the
shear modulus is more delicate and is not available to the best of our knowledge.}. 
We take as an upperbound
of the change in strain
$\delta \epsilon$ one-third the relative change in density ($11~\%$). 
The volume $V$ is the surface
$S$ of the fault times the width $w \approx 0.1~meter$, say,
of its core over which the transformation 
occurs. We thus see that this corresponds to the seismic moment of an earthquake
with average slip $w~\delta \epsilon$ less than a centimeter, irrespective of the 
size of the fault. This contribution is thus negligible for large earthquakes having
slips of meters or more. In addition to this double-couple component, the change in
bulk modulus and in density radiates isotropically with radial directions 
for the first $P$ motions and no first $S$ motions. However, this is again a very small effect.

The second contribution to seismic radiation comes from the mechanical fault slip.
Many models have been studied and the seismic radiation efficiency is a function 
of the acceleration and deceleration of the slip.
Modelling this phenomenon of fault unlocking
 is difficult due to the fact that friction is a complex function of slip, velocity and
 other parameters. One thus has to solve a nonlinear problem
of wave propagation and radiation with moving boundary conditions (Stefan's problem).

Another essentially unsolved problem is the stopping of the slip or
the arrest of the rupture. The seismic radiation and the resulting post-seismic
stress are controlled by the nature of the arrest. Does it occur when the 
sliding velocity goes to zero as in standard solid friction or abruptly by a sudden
collision with a barrier? We do not know. Within the acoustic fluidization mechanism, 
the sliding motion lasts as long as the 
high-frequency waves which are trapped within the liquified material with low-acoustic impedence 
remain of sufficient amplitude to unlock the fault. The detrapping of these waves control the
healing of the fault and therefore the static stress drop. This mechanism
is expected to produce large variations of static stress drops, depending on
the degree of chemical alteration and the detailed
mineralogy in the fault core and the availability of fluids.

\subsection{Earthquake dynamical rupture}

In this section, we study a simplified case in which 
a crack suddenly appears within the
crust and leads to a given stress drop $\Delta \sigma$. This exemple is not
the most common view according to which a rupture front propagates
from the nucleation center to the edges of the fault. It however allows us
to present the salient features of elasto-dynamic effects. Physically,
it could correspond to the hypothetical case where another mechanism, such as a shock or
and electromagnetic impulse would propagate much faster than the shear sound
velocity and prepare almost simultaneously the fault for rupture. 
It is also of interest because the static version has been examined in details
[{\it Knopoff}, 1958].

In the presence of the tectonic loading, this sudden crack formation creates an
elasto-dynamic transient that relaxes the mechanical system to a new stress and strain
configuration, corresponding to a new equilibrium in the presence of the open crack. The
static version of this problem, in which a crack is introduced in a stressed medium,
has been considered by Knopoff [1958] to evaluate the total energy released by an
earthquake. Here, we are led to examine the dynamical version of this problem for the
smaller portion of the fault where the nucleation occurs.

Backus and Mulcahy [1976a,b] have shown that any mechanically admissible seismic source
should admit a moment tensor representation. A seismic source, i.e. some internal
rupture of the crust, is a non-elastic stress and strain distribution.
Given the moment tensor density $m(r,t)$, the
equivalent body force density derived from this inelastic stress distribution is
\be
f_{\beta}^{(V)}(r,t) = -\partial_{\alpha} m_{\alpha \beta}(r,t)~~.
\label{forcemoment}
\ee
The divergence expression ensures that the integration of (\ref{forcemoment}) over an
arbitrary volume with no geometrical or stress discontinuity gives zero, which is a
first requirement for the body to be at equilibrium. A second requirement is that the
total torque in the same volume be zero which leads to the property that $m(r,t)$ is
a symmetric tensor. It has thus all the properties of a stress tensor. The integration
of (\ref{forcemoment}) over a volume with one part in the solid and one part in the
empty interior of the crack (in which the stress is equal to the 
large scale value minus the stress drop $\Delta \sigma$) leads to the relation
between the surface force and the moment\,:
\be
f_{\beta}^{(S)}(r,t) = n_{\alpha} m_{\alpha \beta}(r,t)~~,
\label{forcemomentdf}
\ee
where $n_{\alpha}$ is the $\alpha$ component of the unit (out-going) vector normal to
the surface.

What is the moment tensor density $m(r,t)$ which is equivalent to the sudden
opening of a crack? The crack is defined uniquely by the geometrical locii of points
on which the stress is zero. By the linearity of elastic equations and the
corresponding principle of superposition, the creation of a stress-free crack is the
same as imposing a singular stress source at the position of the crack and which
opposes locally the stress field $\sigma_{ij}^0$ established globally prior to the
creation of the crack. The sum of the pre-existing stress and the singular stress
gives a stress decreased by the amount of the stress drop 
exactly on the position of the crack. This singular stress
source can be called the stress drop due to the creation of the crack. This
simple reasoning recovers the general result that the moment tensor density is
nothing but the stress drop, when including the surface contributions [{\it Gilbert},
1971; {\it Backus and Mulcahy}, 1976]. We thus get 
\be
m_{ij}(r,t) = - \Delta \sigma_{ij}~ \delta(r-r_c) ~{\cal H}(t)~~,
\label{memhfk}
\ee
where $\delta(r-r_c)$ expresses the fact that the moment tensor source is localized
on the crack position. ${\cal H}(t)$ is the Heaviside's function ($=0$ for $t<0$ and
$=1$ for $t \geq 0$) and $0$ is the origine of time at which the crack is created inside 
the medium.

Using the representation theorem [{\it Aki and Richards}, 1980], we can write the
elastic displacement created by the source (\ref{memhfk}) as [{\it Madariaga}, 1981]
\be
u_i(r,t) = \int_0^t dt_0 \int_{V_0} dV {\partial G_{ij}(r,t|r_0,t_0) \over \partial
x_k} m_{jk}(r_0,t_0)~~,
\label{gfdvcg}
\ee
obtained from quiescent initial conditions $u(r,t) = {\partial u(r,t) \over \partial
t} = 0$ and the elasto-dynamic Green function is [{\it Aki and Richards}, 1980]
$$
 G_{ij}(r,t|r_0,t_0) = {1 \over 4\pi \rho} (3 \gamma_i \gamma_j - \delta_{ij}) 
{t \over R^3} \biggl[ {\cal H}(t-{R \over v_p}) - {\cal H}(t-{R \over v_s}) \biggl]
$$
\be
+ {1 \over 4\pi \rho v_p^2} \gamma_i \gamma_j {1 \over R} \delta (t-{R \over v_p})
+ {1 \over 4\pi \rho v_s^2} (\delta_{ij} - \gamma_i \gamma_j) {1 \over R} 
\delta (t-{R \over v_s})~~.
\label{reyjkks}
\ee
In (\ref{reyjkks}), $\rho$ is the density of the crust, $\gamma_i$ are the cosine
directors of $r-r_0$, $R = |r-r_0|$, $v_p = \sqrt{\lambda + 2 \mu \over \rho}$ and 
$v_s = \sqrt{\mu \over \rho}$ are respectively the P (compressive) and S (shear) wave
velocities. The first term in (\ref{reyjkks}) is the near field term which decays
much faster than the two others which correspond to the classical P and S body waves.

Inserting (\ref{memhfk}) and (\ref{reyjkks}) in (\ref{gfdvcg}), performing the
integrals over the delta in time and in space as well as the derivative with respect
to $x_k$ yields the following expression for the elasto-dynamic displacement due to
the P and S waves
$$
- u_i(r,t) = {\Delta \sigma_{jk} \over 4 \pi \rho} \int_{S_0} dS 
[{\delta_{ij} \over R} - \gamma_i \gamma_j (1-\delta_{ij})]
\biggl[{1 \over
v_p^2}{\cal H}(t - {R \over v_p}) - {1 \over v_s^2}{\cal H}(t - {R \over v_s}) \biggl]
$$
\be
- {\sigma_{jk}^0 \over 4 \pi \rho} \int_{S_0} dS {\gamma_i \gamma_j \gamma_k \over R}
\biggl[{1 \over v_p^3} \delta(t - {R \over v_p})
- {1 \over v_s^3} \delta(t - {R \over v_s}) \biggl]
+ {\sigma_{ik}^0 \over 4 \pi \rho v_s^2} \int_{S_0} dS 
{\gamma_k \over R} \biggl[ {\cal H}(t - {R \over v_s}) - {1 \over v_s} 
\delta (t - {R \over v_s}) \biggl]~~,
\ee
where $S_0$ denotes the crack surface. This is the seismic signal measured
at position $r$ and time $t$.

We can use these expressions to calculate the stress impulse that the border of the
crack will receive and whether it can rupture further due to the stress loading. The
on-going rupture is then that of a fault with friction that has been loaded
dynamically by the local crack rupture. This regime has been analyzed recently in the
litterature.

Freund [1990] provides the complete treatment of the problem of a suddenly applied loading
on a half-plane crack. 
The solution for the antiplane case is given in equation (2.3.16), p. 70 in [{\it Freund}, 1990]
and leads to a shear stress close to the crack tip equal to 
\be
\sigma(r,t) \sim {2 \Delta \sigma \over \pi} \sqrt{ct \over r}~,
\label{cfdfgxgx}
\ee
where $r$ is the distance to the crack tip.
The case of the suddenly applied in-plane shear traction, corresponding in our problem
to the full tensorial elastic case with stress drop $\Delta \sigma$ over the 
whole extension of the crack, is treated in [{\it Freund}, 1990], section 2.6, p. 97. 
The result (\ref{cfdfgxgx}) still holds, with $c$ replaced by the longitudinal acoustic
wave velocity.

\section{Other observations and questions}

We present a series of questions and/or observations, not all firmly established (little
is found non-controversial in this field), but which provide stimulating questions that
an understanding of the earthquake process should embrace.

\subsection{Laboratory experiments}

To our knowledge, laboratory experiments of rock friction have not explored
fully the parameters that are relevant in seismogenic conditions. In particular,
we mentionned already the difficulties in constraining chemical activity. Another
relevant question is the role of possible mineralogic 
transformations of crystalline structures at
conditions of temperature, pressure 
and water content, in the presence of deformation,
that would be relevant for shallow seismogenic depths.
With respect to the dependence with pressure, recent simulations [{\it Badro et al.}, 1996]
suggest to explore also the relevance of nonhydrostatic compression, for instance in 
triaxial mechanical set-ups.

\subsection{The longer the recurrence time, the larger the stress drop}

The average slip rate of faults varies very significantly from a small fraction 
of $1~mm/year$ to several $cm/year$.
Several studies have shown recently a remarkable relationship between the average
slip rate on faults and the stress drop associated to earthquakes occurring on these
faults (see {\it Scholz}, 1990, p.408; {\it Kanamori}, 1994) for reviews). What is obvious 
(within the standard rebound earthquake model) is that
 faults with slow (resp. fast) slip rates usually have long (resp. short) repeat times.
 A number of works
 ({\it Kanamori and Allen}, 1986; {\it Scholz et al.}, 1986; {\it Houston}, 1990) have found
 that earthquakes on faults with long repeat times (thousands of years) 
 radiate more energy per unit fault
 length and have a significant 
 larger dynamical stress drop than those with short repeat times (a few decades to
 a few centuries). This is quite puzzling\,: how can the loading process, occurring on a
 very slow time scale of centuries,
  have an influence on the dynamics occurring at the time scale of seconds?
 The usual explanation is that faults heal progressively (by plasticity and
 chemical processes) and thus become stronger as
 more time is given to them between earthquakes. However, we note that this is at variance
 with the other piece of argument usually brought forth, namely that an earthquake
 can occur because of a progressive weakening due to corrosion. Also, a stronger fault should
 produce a larger frictional heating and need a larger stress to break down.

\subsection{Shear heating}

Direct measurements of heat flux do not exhibit anomalous values and thus no
significant frictional heating [{\it Henyey
and Wasserburg}, 1971; {\it Lachenbruch and Sass}, 1980; 1988; 1992; 1995;
{\it Sass et al.}, 1992]. On the other hand, the accumulated weight of evidence
on the metamorphism of rocks adjacent to faults in higher grades 
has been taken as an indication that shear heating can be associated with faulting 
[{\it Scholz}, 1980; {\it Sibson}, 1980; {\it Frischbutter and Hanisch}, 1991; {\it Peacock}, 1992].
Interpreted with a pure mechanical point of view, this implies shear stresses of about
$100~MPa$ at least at the initiation of slip, in contradiction with many other 
interpretations. A mechanism has recently been proposed [{\it Harrison et al.}, 1997] that 
renders unncessary the need for exceptional physical conditions (such as high shear stress), to
explain the generation of the Himalayan leucogranites. In this example, 
the apparent inverted metamorphism appears
to have resulted from activation of a broad shear zone beneath the Himalayan Main Central thrust 
which tectonically telescoped the young metamorphic sequence.

\subsection{Seismic P-wave nucleation phase}

Seismic P-wave signals have been reported [{\it Ellworth and Beroza}, 1995; {\it
Beroza and Ellworth}, 1996; {\it Dodge et al.}, 1996], 
that seem to precede the arrival of the first P-wave
(longitudinal compressive acoustic wave).  These observations are still controversial
[{\it Mori and Kanamori}, 1996], not only because the reported
signals are weak and the effect is hard to establish, 
but also because their presence is essentially ruled out within
the standard picture. Indeed, first, one would have to invent the existence of 
a new mode that propagated faster than the P-wave {\footnote{Guided waves
that have been found to propagate within the core of faults
[{\it Li et al.}, 1994] are slower, by the very constraint that they are trapped.
If they were faster, they would leak immediately\,: this is the usual condition for 
the existence of evanescent waves.}}. Secondly, 
[{\it Ellworth and Beroza}, 1995; {\it
Beroza and Ellworth}, 1996; {\it Dodge et al.}, 1996] make the remarkable claim that
 the duration of the P-wave precursors is roughly proportional to the total length 
of the earthquake rupture\,: it is as if the precursor, supposedly associated to 
a small nucleation zone, knew in advance about the (large) size of the earthquake; or, that
the size of the nucleation zone controlled the size of the large earthquake.
Present researches attempt to understand these observations using the slip dependence of
friction given by the Ruina-Dieterich laws 
[{\it Dieterich}, 1972; 1978; 1979; 1992; {\it Ruina}, 1983].

\subsection{Earthquakes occur preferentially in strong rocks}

Fault imaging shows the systematic picture that earthquake rupture takes
place in rock with relatively high seismic velocity, while surrounding rocks have
velocities that are relatively low. The same pattern emerges in images of the 1994
Northridge earthquake [{\it Zhao and Kanamori}, 1995], of the 1966 Parkfield shock
of 6.0 and of the 1992 Landers earthquake of magnitude 7.3. Why do earthquakes
regularly take place in strong rock? If we consider an heterogeneous fault, which
weak and strong rocks distributed along its length, we do expect on mechanical
grounds that the large earthquakes will nucleate on the strong rocks. Indeed, only
strong asperities can store sufficient stress to produce a rupture that will
not be easily stopped. This scenario has been verified in block-spring models with
heterogeneous friction along the fault [{\it Knopoff}, 1996b]. 
But this does not account for the observation that earthquakes take place 
in rocks that are stronger than the surrounding rocks in the direction {\it transverse}
to the fault.

\subsection{Earth tides and seismicity remotely triggered at long distances}

The long-term tectonic loading stress rate of the order of $10^{-3}~bar/hr$ is much less
than the stress rate up to $0.15~bar/hr$ 
due to earth tides from gravitational interactions with the Moon and Sun.
Tidal triggering of earthquakes would thus be expected if rupture began soon
after the achievement of a critical stress level. Correlations between tides 
and earthquakes have been investigated repeatedly [{\it Emter}, 1997; {\it Cotton}, 1992].
The most careful statistical studies have found no evidence of triggering
[{\it Heaton}, 1982; {\it Rydelek et al.}, 1992; {\it Tsuruoka et al.}, 1995;
{\it Vidale et al.}, 1998]. Two existing theories can explain these observations
by invoking high stress rates just before failure. Dieterich's model of state-
and rate-dependent friction predicts high stressing rates accross earthquake nucleation
zones [{\it Dieterich}, 1997]. Alternatively, changes in fluid plumbing of the fault 
system could conceivably be more rapid than tidal strains and may trigger failure
[{\it Sibson}, 1973].

There is strong evidence that the Landers earthquake, June 28, 1992, southern California,
 triggered seismicity at distances equal to many times the source size [{\it Hill et al.},
1993], with a rate which was maximum immediately after passage of the exciting seismic waves.
The problem is that the dynamical stress created by the seismic waves is very small
at these long distances, of the order of the effect of lunar tides ($0.01~MPa$ and less) which
have not been found to be correlated with earthquakes. Sturtevant et al. [1996] have
recently proposed a model in which the earthquakes are triggered by a rapid increase of
pore pressure due to rectified diffusion of small pre-existing gas bubbles in faults
embedded in hydrothermal systems. This model depends on the confluence of several 
favorable conditions, in particular supersaturated gas in water, large mode conversion
occurring in the geothermal field and a very low ($50~m/s$) shear velocity in the 
porous media filled with the bubbly liquid.

\subsection{Non-double couple components}

Some earthquakes with non-double-couple mechanisms have been claimed
not to be explained solely by a
composite rupture [{\it Frohlich}, 1994b]. The standard explanation for 
non-double-couple components relies on the fault zone irregularity 
[{\it Kuge and Lay}, 1994]. Could they be witnesses of a 
compensated linear dipole component associated with other phenomena
such as structural transformations?

\subsection{Creeping sections and slow earthquakes}

Some faults appear not to rupture in sudden earthquakes but rather to creep
continuously. Slow earthquakes [{\it Linde et al.}, 1996] are acceleration of 
creep over a few hours to a few days. This leads to the usual classification
in terms of three types of behavior\,: regions where the seismic coupling is total\,; those
where slip is seismic only when triggered by seismic slip from an adjacent region
and which otherwise slip asesimically\,; and regions where slip is always
aseismic. Within the standard stick-slip friction theory, these three regimes can be
ascribed to different friction stability conditions, controlled by the 
slip velocity parameter of the friction law which are material properties [{\it
Scholz}, 1990]. 

Consider for instance the famous San Andreas fault system. This long fault
evidently cuts through many different types of rocks and
regional structures along its traverse from the Transverse Ranges to the Mendocino triple junction,
and the patterns of seismicity differ strikingly along various segments of the fault. These
differences in seismic behavior coincide so closely with certain geologic situations along the San
Andreas fault in central California as to suggest a causal relation [{\it Wallace}, 1991].

\subsection{Deep earthquakes}

Solid-solid transformations have been invoked to explain the puzzle of deep-focus
earthquakes.  The paradox is that brittle fracture and frictional sliding at
depths in excess of $100-200 km$ \footnote{The deepest earthquakes occur at depth
$600 km$, which is well within the mantle.} would require unrealistic rock
strength, while the rheology of the rocks becomes more and more ductile at pressure
and temperature increases, as found in laboratory experiments. Furthermore, the
isostatic control of topographic relief implies that the mantle has no strength 
on long time scales at depths greater than $70 km$ or so [{\it Jeffreys}, 1929].
Several mechanisms have been proposed for deep earthquakes [{\it Frohlich}, 1994a],
such as plastic instabilities [{\it Hobbs and Ord}, 1988], shear-induced melting
[{\it Post}, 1977], instabilities accompanying recrystallization [{\it Ogawa},
1987], polymorphic phase transformation [{\it Kirby}, 1987] and olivine$\to$spinel
anticracks formation [{\it Green and Burnley}, 1989]. Each of these proposed
mechanisms exhibits certain weakness [{\it Kirby}, 1987]. While a consensus has
not yet emerged, one of the most prefered mechanism involves so-called {\it
transformational faulting} in which metastable olivine transforms under stress to
denser spinel in thin shear zones along planes of maximum shear stress. Then, it
is proposed that newly formed, fine-grained spinel is the `fluid lubricant' and/or
the damage (through the operation of `anticracks') that makes catastrophic
faulting possible.

A recent analysis of the deep Bolivian earthquake of 9 June 1994 has been proposed
to be incompatible with an origin by transformational faulting [{\it Silver et
al.}, 1995], because the rupture extended to more than three times that expected
for a metastable wedge of olivine at the core of a subducting slab. These authors
suggest that failure during deep-focus earthquakes represent the reactivation of
faults that were established when the lithosphere was near the earth surface.

\subsection{Precursory phenomena}

Since the inception of seismology, the search for reliable precursory phenomena
of earthquakes has been very active.
Current research suggests that the seismic process is preceded by a complex
set of physical precursors. However, there is no concensus on the 
statistical significance of these precursors and their reliability.
The lack of agreement among prediction advocates on a single set of
``best-candidate precursors'' underscores the weakness and inconsistency of
their case [{\it Wyss}, 1991, 1997]. To validate a prediction
method, one must show it to be successful beyond random chance, but
such success has not been demonstrated convincingly. The problem is very difficult
because of the complexity of earthquake phenomenology and the low rate occurrence of 
earthquakes.

Seismic activity is one obvious candidate. It 
is monitored in the different parts of the world by
arrays of seismometers which allow for the determination of the epicenter,
hypocenter, and rupture parameters (magnitude, seismic moment, source
mechanism, orientation of the fault plane and direction of motion) of an
earthquake. Foreshocks correspond to an increase of seismic activity at time scales of
no more than a few weeks before a ``main'' shock [{\it Abercrombie and Mori}, 1996;
{\it Maeda}, 1996; {\it Jones}, 1994]. Only about $30\%$ of earthquakes
exhibit foreshocks and the problem is that they are recognized as such after the fact.
Nevertheless, foreshocks as well as aftershocks indicate the existence of a delay
mechanism [{\it Shaw}, 1993] in the activation of earthquakes which could possibly be used
[{\it Sornette et al.}, 1992]. 

The notion of foreshocks has been extended to that of seismic precursors to 
describe the surprising and somewhat controversial observations
showing that many large earthquakes have been
preceded by an increase in the number of intermediate sized events 
[{\it Jaum\'e and Sykes}, 1998]. The
relation between these intermediate sized events and the subsequent main
event has only recently been recognized because the precursory events occur
over such a large area that they do not fit prior definitions of foreshocks.
In particular, the eleven
earthquakes in California with magnitudes greater than $6.8$ in the last
century are associated with an increase of precursory intermediate magnitude
earthquakes measured in a running time window of five years [{\it Knopoff et al.}, 1996].
What is
strange about the result is that the precursory pattern occured with
distances of the order of $300$ to $500 ~km$ from the futur epicenter, 
i.e. at distances up to ten times larger that the size of the futur earthquake rupture. 
Furthermore, the increased intermediate magnitude activity switched
off rapidly after a big earthquake in about half of the cases. This implies
that stress changes due to an earthquake of rupture dimension as small as
$35 ~km$ can influence the stress distribution to distances more than ten
times its size. This result defies usual models.

This observation is not isolated. There is mounting evidence, albeit still controversial,
that the rate of occurrence of intermediate earthquakes increases in the
tens of years preceding a major event. Sykes and Jaume [1990]
present evidence that the occurrence of events in the range $5.0-5.9$
accelerated in the tens of years preceding the large San Francisco bay area
quakes in 1989, 1906, and 1868 and the Desert Hot Springs earthquake in
1948. Lindh [1990] points out references to similar increases in
intermediate seismicity before the large 1857 earthquake in southern
California and before the 1707 Kwanto and the 1923 Tokyo earthquakes in
Japan. More recently, Jones [1994]  has documented a
similar increase in intermediate activity over the past $8$ years in southern
California. This increase in activity is limited
to events in excess of $M = 5.0$; no increase in activity is apparent when
all events $M > 4.0$ are considered. Ellsworth et al. [1981] also
reported that the increase in activity was also limited to events larger
than $M = 5$ prior to the 1989 Loma Prieta earthquake in the San Francisco
Bay area. Bufe and Varnes [1993] analysed the increase
in activity which preceded the 1989 Loma Prieta earthquake in the San
Francisco Bay area while Bufe et al. [1994] document a current increase in
seismicity in several segments of the aleutian arc. More recently, Bowman et al. [1998]
conducted a systematic review of seismicity in California specifically aimed to test
for the occurrence of accelerating seismic strain release before large earthquakes.
They examined all $M \geq 6.5$ earthquakes along the San Andreas fault system since 1950
(total of eight). They found that the cumulative seismic strain prior to all eight 
$M \geq 6.5$ earthquakes can be better fit by a power acceleration than by a steady seismic
rate. Brehm and Braile [1998] has used the same technique as Bowman et al. [1998]
to study earthquakes in the New Madrid Seismic Zone in the central United States. Surprinsingly,
they find they can model accelerating seismic strain sequences before mainshocks as small
as magnitude $3.5$ that yield reasonable predictions of time and magnitude\,: of twenty-six
post-1979 mainshocks ranging from $3.5$ to $5.2$, sixteen could be modelled by a
power law accelerating time-to-failure function.

Seismic activity induces other
variations of geophysical parameters such as the surrounding electric and
magnetic fields. Under favorable conditions these electromagnetic
variations may be observed before seismic rupture. 
A recent workshop on earthquake precursors held at The University of
Electro-Communications, Tokyo, and sponsored by The National Space
Development Agency of Japan, illustrated a wide range of
seismo-electromagnetic precursor radiation [{\it NASDA-CON-960003}, 1997]. 
The ability of the Earth's crust in areas of active
deformation to transform mechanical energy into electromagnetic waves is a
possible contribution to the short-term prediction of seismic hazards.
 The most important and difficult problem in seismo-electromagnetic research
is the location of source. This problem is related to both the physical
processes in the earthquake preparation zone and the choice of monitoring
device.  In a re-examination of the VAN precursors [{\it Varotsos and Lazaridou}, 1991],
Mulargia and Gasperini [1992] suggested that electric signals are more readily found
to follow rather than precede large events. 

Other potential signatures of impending earthquakes involve
the behavior of global quantities, such as the conductivity and
permeability of the crust at various distances from the epicenter.
The change in permeability, controlled by microcracking and a change in porosity, 
can be monitored through the fluctuations of groundwater. In this spirit,
fluctuations in ion concentrations of groundwater issuing from
wells located near the epicenter of the recent earthquake of
magnitude 6.9 near Kobe, Japan, on January 17, 1995 
have been observed [{\it Tsunogai and Wakita}, 1995, 1996]
and quantified [{\it Johansen et al.}, 1996].

\section{Concluding remarks}

This review has attempted to bring together different disciplines that
have hitherto been developed in large part independently from each other, while 
they study fundamentally the same ``object''. We thus view
the challenge of understanding what is an earthquake as somewhat parallel to the integration
of many disparate observations in seismicity, volcanism, mountain-building and
paleo-magnetism which were finally made sense of and 
unified within the theory of plate tectonics.

The review has been built on 
the difficulties to reconcile data with existing models and on the recognition of the
fundamental role played by water in fault zone.
The possible scenarios for an earthquake that emerge
are not simple and invoke a series of mechanisms, which are
naturally interwoven. This may suggest directions for 
a coherent framework to explain a large set of diverse
observations in earthquakes. It also raises the exciting promise of providing a
guideline for recognizable and reproducible precursors that could be exploited. 

\vskip 1cm

{\bf Acknowledgments}: I have benefitted from very helpful discussions with Y. Ben-Zion,
Y. Brechet,
J.M. Christie, P.M. Davis, J.Dieterich, P. Evesque, J. Gilman, M. Harrison, H. Houston, D.D. Jackson, 
Y.Y. Kagan, W. D. Ortlepp, G. Ouillon, E. Riggs and J. Vidale.  
 Useful correspondences with J.-C. Doukhan, J. Ferguson, R. J. Geller, I. Main, 
 G. Martin, J.P. Poirier
are acknowledged. Special thanks are due to L. Knopoff and A. Sornette for many discussions and
interactions over many years.

\pagebreak

\section{References}
{\small

Abercrombie, R.E. and J. Mori, Occurrence patterns of foreshocks to large earthquakes in
the western United States, {\it Nature, 381}, 303-307, 1996.

Aines, R.D., and G.R. Rossman, Water in minerals? A peak in the infrared, 
{\it J.Geophys.Res., 89}, 4059-4071, 1984.

Aki, K., Characterization of barriers on an earthquake fault, 
{\it J.Geophys.Res., 84}, 6140-6148, 1979.

Aki, K., Impact of earthquake seismology on the geological community since the
Benioff zone, {\it Geol. Soc. Am. Bull., 100}, 625-629, 1988.

Aki, K., and P.G. Richards, {\it Quantitative seismology}, Freeman, San Francisco,
1980.

Alekseevskaya, M.,  A. Gabrielov, I. Gel'fand, A. Gvishrani and E. Rantsman, 
{\it J. Geophys. 3}, 227, 1977.

Anderson, E.M., The dynamics of faulting, 2nd ed., (Oliver and Boyd, Edinburgh, 1951).

Anderson, P.W., More is different, {\it Science, 177}, 393-396, 1972.

Andrews, D.J., Mechanics of fault junctions, {\it J.Geophys.Res., 94},
9389-9397, 1989.

Anooshehpoor, A., and Brune J.N., Frictional heat generation and seismic
radiation in a foam rubber model of earthquakes, 
{\it Pure and Applied Geophysics, 142},735-747, 1994. 

Atkinson, B.K. et al., Mechanisms of fracture and friction of crustal rock in
simulated geologic environments,  supported by the Earthquake
Hazards Reduction Program, [Menlo Park, Calif.] : U.S. Geological Survey, 81-277,
1981.

Atkinson, B.K., Subcritical crack growth in geological materials,
{\it J.Geophys.Res., 89}, 4077-4114, 1984. 

Backus, G. and M. Mulcahy, Moment tensors and other phenomenological descriptions of
seismic sources -- I Continuous displacements, {\it Geophys. J. R. astr. Soc., 46},
341-361, 1976a.

Backus, G. and M. Mulcahy, Moment tensors and other phenomenological descriptions of
seismic sources -- II Discontinuous displacements, {\it Geophys. J. R. astr. Soc.,
47}, 301-329, 1976b.

Badro, J., J.-L. Barrat and P. Gillet, Numerical simulation of $\alpha$-quartz
under nonhydrostatic compression\,: memory glass and five-coordinated crystalline
phases, {\it Phys. Rev. Lett., 76}, 772-775, 1996.

Baeta, R.D. and K.H.G Ashbee, Slip systems in quartz, I-Experiments;
II-Interpretation, {\it Am. Mineral., 54}, 1551-1582, 1970.

Bailey, G., G.C.P. King and D. Sturdy, Active tectonics and land-use strategies -
A paleolithic example from northwest Greece, {\it Antiquity, 67}, 
292-312, 1993.

Bak, P., and C. Tang, Earthquakes as self-organised critical phenomenon, 
{\it J. Geophys. Res. 94}, 15635-15637, 1989.

Bambauer, H.U., G.O. Brunner and F. Laves, Light scattering of heat-treated quartz
in relation to hydrogen containing defects, {\it Am. Mineral., 54}, 718-724, 1969.

Bardet, J.P., Damage at distance, {\it Nature, 346}, 799, 1990.

Barton, C.A. and M.D. Zoback, Stress perturbations associated with active faults
penetrated by boreholes - Possible evidence for near-complete stress drop and a
new technique for stress magnitude measurement, 
{\it J. Geophys. Res. 99}, 9373-9390, 1994.

Baumberger, T., and L. Gauthier, Creeplike relaxation at the interface 
between rough solids under shear, {\it J. Phys. I France, 6}, 1021-1030, 1996.

Beeler, N.M., T.E. Tullis, M.L. Blanpied and J.D. Weeks, 
Frictional behavior of large displacement experimental faults,
{\it J. Geophys. Res., 101}, 8697-8715, 1996.

Beeler, N.M., T.E. Tullis and J.D. Weeks, 
The roles of time and displacement in the evolution effect in rock
friction, {\it Geophys. Res. Lett., 21}, 1987-1990, 1994.

Beeman, M., W.B. Durham and S.H. Kirby, Friction of ice, 
{\it J. Geophys. Res., 93}, 7625-7633, 1988.

Behrmann, J.H. and D. Mainprice, Deformation mechanisms in a high-temperature
quartz-feldpar mylonite\,: evidence for superplastic flow in the lower continental
crust, {\it Tectonophysics,140}, 297-305, 1987.

Benzion, Y., and Rice J., Earthquake failure sequences along a cellular fault zone in
a 3-dimensional elastic solid containing asperity and nonasperity regions, 
{\it J. Geophys. Res. 93}, 14109-14131, 1993.

Benzion, Y., and Rice J., Slip patterns and earthquake populations
along different classes of faults in elastic solids, {\it J. Geophys. Res.100},
12959-12983, 1995.

Beroza, G.C., and W.L. Ellworth,  Properties of the seismic nucleation phase, {\it
Tectonophysics, 261}, 209-227, 1996.

Beyer, W. H., editor, {\it CRC standard mathematical tables and formulae},
29th ed.,  Boca Raton : CRC Press, 1991.

Biarez, J., and P.-Y. Hicher, Elementary mechanics of soil behavior\,: saturated
remoulded soils, Rotterdam ; Brookfield, VT : A.A. Balkema, 1994.

Bird, P., New finite element techniques for modeling deformation histories of continents
with stratified temperature-dependent rheology, {\it J. Geophys. Res. 94}, 3967-3990, 1989.

Bird, P., and X. Kong, Computer simulations of California tectonics confirm very low
strength of major faults, {\it Geological Society of Americal Bulletin 106}, 159-174, 1994.

Blacic, J.D., and J.M. Christie, Plasticity and hydrolytic weakening of quartz
single crystals, {\it J. Geophys. Res. 89}, 4223-4239, 1984.

Blanpied, M.L., D.A. Lockner and J.D. Byerlee, An earthquake mechanism based on
rapid sealing of faults, {\it Nature, 358}, 574-576, 1992.

Blanpied, M.L., D.A. Lockner and J.D. Byerlee, Frictional slip of granite at
hydrothermal conditions, {\it J. Geophys. Res., 100}, 13045-13064, 1995.

Bocquet, L., and H.J. Jensen, Phenomenological study of hysteresis 
in quasistatic friction, {\it J. Physique I, 7}, 1603-1625, 1997.

Bolt, B.A., Earthquakes, W.H. Freeman and Co., New York, 1993.

Bowden, F.P., and D. Tabor, The Friction and Lubrication of Solids
(Oxford University Press, 1954).

Bowman, D.D., G. Ouillon, C.G. Sammis, A. Sornette and D. Sornette,
An Observational Test of the Critical Earthquake Concept, {\it J. Geophys. Res.}, in press, 1998.

Brace, W.F., Orientation of anisotropic minerals in a stress field\,: discussion,
{\it Geol. Soc. Am. Memoir, 79}, 9-20, 1957.

Brace, W.F., Laboratory studies of stick-slip and their application to
earthquakes, {\it Tectonophysics, 14}, 189-200, 1972.

Brace, W.F., Permeability of crustalline and argillaceous rocks\,: status and
problems, {\it International Journal of Rock Mechanics in Mineral Sciences and
Geomechanical Abstracts, 17}, 876-893, 1980.

Brace, W. R., and J.D. Byerlee,  Stick-slip as a mechanism for earthquakes,
{\it Science, 153}, N3739, 990-992, 1966.

Bratkovsky, A.M., E.K.H. Salje, S.C. Marais and V. Heine, Strain coupling as the
dominant interaction in structural phase transition, {\it Phase transitions, 55},
79-126, 1995.

Brehm, D.J. and L.W. Braile, Intermediate-term prediction using precursory events
in the New Madrid Seismic Zone, {\it Bull. Seism. Soc. Am., 88}, 564-580, 1998.

Brodie, K.H., Variation in amphibolite and plagioclase composition with
deformation, {\it Tectonophysics,78}, 385-402, 1981.

Bruhn, R.L., W.T. Parry, W.A. Yonkee and T. Thompson, Fracturing and hydrothermal
alteration in normal fault zones, {\it Pure Appl. Geophys., 142}, 609-644, 1994.

Brune, J.N., S. Brown and P.A. Johnson, Rupture mechanism and interface separation
in foam rubber models of earthquakes\,: a possible solution to the heat flow
paradox and the paradox of large overthrusts, {\it Tectonophysics, 218}, 59-67,
1993.

Bufe, C.G. and D.J. Varnes, Predictive modeling of the seismic cycle of the 
greater San-Francisco bay region, {\it J. Geophys. Res., 98}, 9871-9883, 1993.

Bufe, C.G., S.P. Nishenko and D.J. Varnes, Seismicity trends and potential for large
earthquakes in the Alaska-Aleutian region, 
{\it Pure and Applied Geophysics, 142}, 83-99, 1994.

Bullard, E.C., The interior of the earth, pp.57-137 in ''the Earth as a
planet'', G.P. Kuiper, ed., University of Chicago Press, 1954 (see pp. 120-121).

Burridge, R. and L. Knopoff, Body force equivalents for seismic dislocation, {\it
Seism. Soc. Am. Bull., 54}, 1875-1888, 1964.

Burst, J.P., Diagenesis of Gulf Coast clayey sediments and its possible relation
to petroleum migration, {\it American Association of Petroleum Geologists
Bulletin, 53}, 73-79, 1969.

Byerlee, J., Friction of rocks, In Experimental studies of rock friction with
application to earthquake prediction, ed. J.F. Evernden, U.S. Geological Survey,
Menlo Park, Ca, 55-77, 1977.

Byerlee, J., Friction, overpressure and fault normal compression, {\it Geophys. Res.
Lett., 17}, 2109-2112, 1990.

Byerlee, J. D., et al., Stick slip, stable sliding, and
earthquakes - Effect of rock type, pressure, strain rate, and stiffness,
{\it J. Geophys. Res., 73}, 6031-6037, 1968.

Campillo, M., J.C. Gariel, K. Aki and F.J. Sanchez-Sesma, Destructive strong ground
motion in Mexico-city -- Source, path and site effects during 1985 Michoacan
earthquake, {\it Bull. Seism. Soc. Am., 79}, 1718-1735, 1989.
   
Carlson, J.M., J.S. Langer and B.E. Shaw, Dynamics of earthquake faults, 
{\it Rev. Mod. Phys. 66}, 657-670, 1994.

Caroli, C., and P. Nozi\`eres,  Dry friction as a hysteretic elastic response,
In Physics of Sliding Friction, Persson et Tosatti, eds., Kluwer
Academic Publishers, 1996.

Caroli, C., and B. Velicky,
Dry friction as an elasto-plastic response: Effect of compressive
plasticity, {\it J. Physique I, 7}, 1391-1416, 1997.

Carter, N.L., A.K. Kronenberg, J.V. Ross and D.V. Wiltschko, Control of fluids on
deformation of rocks, in Knipe, R.J., and E.H. Rutter, eds., Deformation
mechanisms, rheology and tectonics, {\it Geological Society Special Publication,
54}, 1-13, 1990.

Chen, K., Bak, P., and Okubov, S.P., Self-organized criticality in
crack-propagation model of earthquakes, {\it Phys. Rev. A 43}, 625-630, 1991.

Chester, F.M., A rheologic model for wet crust applied to strike-slip faults, {\it J.
Geophys. Res., 100}, 13033-13044, 1995.

Chester, F.M., J.P. Evans and R.L. Biegel, Internal structure and weakening
mechanisms of the San Andreas fault, {\it J. Geophys. Res., 98}, 771-786,
1993.

Chirone, R., L. Massimilla and S. Russo, Bubble-free fluidization of a cohesive
powder in an acoustic field, {\it Chemical Engineering Science, 48}, 41-52, 1995.

Chopin, C., Coesite and pure pyrope in high-grade blueschists of the Western Alps\,:
a first record and some consequences, {\it Contributions to Mineralogy and Petrology,
86}, 107-118, 1984.

Choy, G.L., and J.L. Boatwright, Global patterns of radiated seismic energy and
apparent stress, {\it J. Geophys. Res., 100}, 18205-18228, 1995.

Cochard, A., and R. Madariaga, Dynamic faulting under rate-dependent
friction, {\it Pure and Applied Geophysics 142}, 419-445 (1994).

Cochard, A., and R. Madariaga, Complexity of seismicity due to highly rate-dependent
friction, {\it J. Geophys. Res. 101}, 25321-25336, 1996.

Cohen, J.E., and P. Horowitz, Paradoxical behaviour of mechanical and electrical
networks, {\it Nature, 352}, 699-701, 1991.

Cotton, L.A., Earthquake frequency, with special reference to tidal stresses in the
lithosphere, {\it Bull. Seism. Soc. Am., 12}, 47-198, 1992.

Courant, R., and K.O. Friedrichs, Supersonic flow and shock waves,
Springer-Verlag, New York and Berlin, third edition, 1985.

Cowie P., Vanneste C. and Sornette D., Statistical physics model of complex fault
pattern organization and earthquake dynamics, {\it J.Geophys.Res. 98}, 21809-21821,
1993.

Cox, S.J.D., Velocity-dependent friction in a large direct shear experiment on
gabbro, In Deformation Mechanisms, Rheology and Tectonics, Knipe et Rutter, eds., 
vol. 54, 63-70, Geological Society Special Publication, 1990.

Christian, J. W., {\it Theory of transformations in metals and alloys\,; an advanced
text book in physical metallurgy}, Oxford, New York, Pergamon Press, 1965.

Dahl, P.S., and M.J. Dorais, Influence of $F(OH)_{-1}$ substitution on the
relative mechanical strength of rock-forming micas, {\it J. Geophys. Res., 101},
11519-11524, 1996.

Darling, R.S., I.M. Chou and R.J. Bodnar, An occurrence of metastable cristobalite in
high-pressure garnet granulite, {\it Science, 276}, 91-93, 1997.

David, C., T.-F. Wong, W. Zhu and J. Zhang, Laboratory measurements of
compaction-induced permeability change in porous rocks\,: implications for the
generation and maintenance of pore pressure excess in the crust, {\it Pure Appl. Geophys.,
143}, 425-456, 1994.

Debate on VAN, Special issue of {\it Geophys. Res. Lett., 23}, 1291-1452, 1996.

Deer, W. A., R. A. Howie and J. Zussman, An introduction to the rock forming
minerals, 2nd ed.,  Harlow, Essex, England\,: Longman Scientific \& Technical\,;
New York, NY, Wiley, 1992.

DeVries, R.C., Diamonds from warm water, {\it Nature, 385}, 485-485, 1997.

Dieterich, J.H., Time-dependent friction in rocks,
{\it J. Geophys. Res., 77}, 3690-3697, 1972.

Dieterich, J.H., Time-dependent friction and the mechanics of stick-slip,
{\it Pure and Applied Geophysics, 116}, 790-806, 1978.

Dieterich, J.H., Modeling of rock friction: 1. experimental results and
constitutive equations, {\it J. Geophys. Res., 84}, 2161-2168, 1979.

Dieterich, J.H., Nucleation and triggering of earthquake slip; effect of periodic
stresses, {\it Tectonophysics, 144}, 127-139, 1987.

Dieterich, J.H., Earthquake nucleation on faults with rate-dependent
and state- dependent strength, {\it Tectonophysics, 211}, 115-134, 1992. 

Dieterich, J., A constitutive law for rate of earthquake
production and its application to earthquake clustering, {\it J. Geophys. Res. 99},
2601-2618, 1993.

Dieterich J. and Kilgore B.D., Direct observation of frictional
constacts- New insight for state-dependent properties, {\it Pure and Applied Geophysics
143}, 283-302, 1994.

Dieterich, J., A constitutive law for rate of
earthquake production and its application to earthquake clustering, {\it J. Geophys.
Res., 99}, 2601-2618, 1993.

Dieterich, J., and Kilgore B.D., Direct observation of frictional
constacts- New insight for state-dependent properties, {\it Pure and Applied
Geophysics, 143}, 283-302, 1994.

Dieterich, J., and Kilgore B.D., Imaging surface contacts -- Power law contact
distributions and contact stresses in quartz, calcite, glass and acrylic plastic, 
{\it Tectonics, 256}, 219-239, 1996.

Dodge, D.A., G.C. Beroza and W.L. Ellsworth, Detailed observations of California
foreshock sequences\,: implications for the earthquake initiation process,
{\it J. Geophys. Res., 101}, 22371-22392, 1996.

Ellsworth, W.L., A.G. Lindh, W.H. Prescott and D.J. Herd,  The 1906
San Francisco Earthquake and the seismic cycle, {\it Am. Geophys.
Union, Maurice Ewing Monogr. 4}, 126-140, 1981.

Ellworth, W.L., and G.C. Beroza, Seismic evidence for a seimic nucleation phase,
{\it Science, 268}, 851-855, 1995.

Emter, D., Tidal triggering of earthquakes and volcanic events, pp. 293-310 of Lecture
Notes in Earth Science 66: Tidal Phenomena (H. Wilhelm, W. Z\"rn and H.-G. 
Wenzel, eds.), Springer-Verlag, Berlin, 1997.

Etheridge, M.A. and J.C. Wilkie, An assessment of dynamically recrystallized
grainsize as a palaeopiezometer in quartz-bearing mylonite zones, 
{\it Tectonophysics, 78}, 475-508, 1981.

Evans, J.P., Textures, deformation mechanisms, and the role of fluids in the
cataclastic deformation of granitic rocks, in Knipe, R.J., and E.H. Rutter, eds.,
Deformation mechanisms, rheology and tectonics, {\it Geological Society Special
Publication, 54}, 29-39, 1990.

Evans, R., Assessment of schemes for earthquake prediction: Editor's introduction, 
{\it Geophys. J. Int., 131}, 413-420, 1997.

Evans, J.P., and F.M. Chester, Fluid-rock interaction in faults of the
San Andreas system - Inferences from San Gabriel fault rock geochemistry
and microstructures, {\it J. Geophys. Res., 100}, 13007-13020, 1995.

Fisher, D.S., K. Dahmen, S. Ramanathan and Y. Ben-Zion,
Statistics of earthquakes in simple models of heterogeneous faults, 
{\it Phys. Rev. Lett. 78}, 4885-4888, 1997.

Fleischmann, K.H., Shallow fluid pressure transients caused by seismogenic normal
faults, {\it Geophys. Res. Lett., 20}, 2163-2166, 1993.

Freiman, S.W., Effects of chemical environments on slow crack growth in glasses
and ceramics,  {\it J. Geophys. Res., 89}, 4072-4076, 1984.

Franssen, R.C.M.W., and C.J. Spiers, Deformation of polycrystalline salt in
compression and in shear at $250-350 ~^{\circ}C$, in Knipe, R.J., and E.H. Rutter, eds.,
Deformation mechanisms, rheology and tectonics, {\it Geological Society Special
Publication, 54}, 201-213, 1990.

Freund, L.B., Dynamic fracture mechanics, Cambridge University Press, Cambridge (U.K.), 1990.

Frischbutter, A., and M. Hanisch, A model of granitic melt formation by
frictional heating on shear planes, {\it Tectonophysics 194}, 1-11, 1991.

Frohlich, C., Earthquakes with non-double-couple mechanisms, {\it Science, 264},
804-809, 1994a.

Frohlich, C., A break in the deep, {\it Nature, 368}, 100-101, 1994b.

Gabrielov, A., V. Keilis-Borok and D.D. Jackson, Geometric incompatibility in a 
fault system, {\it Proc. Nat. Acad. Sci., US 93}, 3838-3842, 1996.

Gavrilenko, P., and Y. Gueguen, Fluid overpressure and pressure solution in the
crust, {\it Tectonophysics, 217}, 91-110, 1993.

Geller, R.J., D.D. Jackson, Y.Y. Kagan and P. Mulargia, Earthquakes cannot be
predicted, {\it Science, 275}, 1616-1617, 1997.

Gilbert, J.F., Excitation of the normal modes of the earth by earthquake sources,
{\it Geophys. J. astr. Soc., 22}, 223-226, 1971.

Goguel, J., Etude m\'ecanique des d\'eformations g\'eologiques, {Bureau de
Recherches G\'eologiques et Mini\`eres, Manuels et M\'ethodes, 6}, 1983.

Gokhberg, M.B., V.A. Morgounov O.A. and Pokhotelov,
Earthquake Prediction: Seismo-Electromagnetic Phenomena. Gordon and Breach
Publishers, Russia, 1995.

Grasso, J.R. and D. Sornette, Testing self-organized criticality by induced
seismicity, {\it J. Geophys. Res.}, in press, 1998.

Gratier, J.P., T. Chen and R. Hellmann, Pressure solution as a mechanism for crack
sealing around faults, natural and experimental evidence, preprint

Green II, H.W., Metastable growth of coesite in highly strained quartz, {\it J.
Geophys. Res., 77}, 2478-2482, 1972.

Green II, H.W., D.J. Griggs and J.M. Christie, Syntectonic and annealing recrystallization
of fine-grained quartz aggregate, in {\it Experimental and Natural Rock Deformation}, 
edited by P. Paulitsch, pp. 272-335, Springer, New York, 1970.

Green II, H.W., and P.C. Burnley, A new self-organizing mechanism for deep-focus
earthquakes, {\it Nature, 341}, 733-737, 1989.

Griggs, D.T., M.S. Patterson, H.C. Heard and F.J. Turner, Annealing
recrystallization in calcite crystals and aggregates, 
{\it Geol. Soc. Am. Memoir, 79}, 21-37, 1957.

Griggs, D.T., F.J. Turner and H.C. Heard, Deformation of rocks at $500~^{\circ}C$ and
$800~^{\circ}C$, {\it Geol. Soc. Am. Memoir, 79}, 39-104, 1957.

Griggs, D.T., and J. Handin, editors, Rock deformation (A symposium), {\it Geol.
Soc. Am. Memoir, 79}, 381 pages, 1960.

Griggs, D.T., and J.D. Blacic, Quartz\,: anomalous weakness of synthetic crystals,
{\it Science, 147}, 292-295, 1965.

Grinfeld, M.A., Instability of the separation boundary between a
nonhydrostatically stressed elastic body and a melt, {\it Sov. Phys. Dokl. 31},
831-834, 1986.

Gu, Y., and T.-F. Wong, Effects of loading velocity, stiffness, and inertia on the
dynamics of a single degree of freedom spring-slider system,
{\it J. Geophys. Res., 96}, 21677-21691, 1991.

Harrison, T.M., F.J. Ryerson, P. Le Fort, A. Yin, O.M. Lovera and E.J. Catlos, 
A late Miocene-Pliocene origin for the central Himalayan inverted metamorphism,
{\it Earth and Planetary Science Letters 146}, E1-E7, 1997.

Hadizadeh, J., Interaction of cataclasis and pressure solution in a
low-temperature carbonate shear zone, {\it Pure Appl. Geophys., 143}, 255-280, 1994.

Heaton, T.H., Tidal triggering of earthquakes, {\it Bull. Seism. Soc. Am., 72}, 2181-2200, 1982.

Heggie, M.I., Molecular diffusion of oxygen and water in amorphous silica\,: role of
basal dislocation, {\it Philos. Mag. Lett., 65}, 155-158, 1992.

Heggie, M.I., A molecular water pump in quartz dislocations, {\it Nature, 355},
337-339, 1992.

Heggie, M.I., R. Jones, C.D. Latham, S.C.P. Maynard and P. Tole, Molecular diffusion
of oxygen and water in crystalline and amorphous silica, {\it Philos. Mag. B, 65},
463-471, 1992.

Heine, V., X. Chen, S. Dattagupta, M.T. Dove, A. Evans, A.P. Giddy, S. Marais, S.
Padlewski, E. Salje and F.S. Tautz, Landau theory revisited, {\it Ferroelectrics,
128}, 255-264, 1992.

Henyey, T.L., and G.J. Wasserburg, Heat flow near major strike-slip
faults in California, {\it J. Geophys. Res., 76}, 7924-7946, 1971.

Herrmann, H.J., G. Mantica and D. Bessis, Space-filling bearings,
{\it Phys.Rev.Lett., 65}, 3223-3226, 1990.

Heuze, F.E., High-temperature mechanical, physical and thermal properties of granitic
rocks -- a review, {\it Int. J. Rock Mech. Sci. \& Geomech. Abstr., 20}, 3-10, 1983.

Hickman, S.H., Stress in the lithosphere and the strength of active faults, {\it
Rev. Geophys., 29}, 759-775, 1991.

Hickman, S., R. Sibson and R. Bruhn, Introduction to special section\,: mechanical
involvement of fluids in faulting, {\it J. Geophys. Res., 100}, 12831-12840, 1995.

Hill, D., P.A. Reasenberg, A. Michael, W.J. Arabaz et al., Seismicity remotely
triggered by the magnitude 7.3 Landers, California, earthquake, {\it Science, 260},
1617-1623, 1993. 

Hobbs, B.E., Recrystallization of single crystals of quartz, {\it Tectonophysics,
6/5}, 353-401, 1968.

Hobbs, B.E., The influence of metamorphic environment upon the deformation of
minerals, {\it Tectonophysics, 78}, 337-382, 1981.

Hobbs, B.E., Point defect chemistry of minerals under a hydrothermal environment,
{\it J. Geophys. Res., 89}, 4026-4038, 1984.

Hobbs, B.E., and A. Ord, Plastic instabilities; implications for the
origin of intermediate and deep focus earthquakes,
{\it J. Geophys. Res., 93}, 10521-10540, 1988.

Houston, H., A comparison of broadband source spectra, seismic energies and stress
drops of the 1989 Loma Prieta and 1988 Armenian earthquakes, {\it Geophys. Res. Lett.
17}, 1413-1416, 1990.

Ishimaru, A., Wave propagation and scattering in random media
(Academic Press, New York, 1978).

Ito K. and M. Matsuzaki, Earthquakes as self-organized critical phenomena, {\it J.
Geophys. Res. B 95}, 6853-6860, 1990.

Janecke, S.U. and J.P. Evans, Feldspar-influenced rock rheologies, {\it Geology,
16}, 1064-1067, 1988. 

Jaum\'e, S.C. and L.R. Sykes, Evolving towards a critical point: a review of
accelerating seismic moment/energy release prior to large and great earthquakes, 
{\it Pure and Applied Geophysics}, submitted, 1998.

Jeffreys, H., The earth, its origin, history and physical constitution, 2nd
edition (Cambridge University Press, Cambridge, UK, 1929).

Jensen, H.J., Y. Brechet and B. Doucot, 
Instabilities of an elastic chain in a random potential,
{\it J. Phys. I France, 3}, 611-623, 1993.

Johansen, A., P. Dinon, C. Ellegaard, J.S. Larsen et H.H. Rugh,
Dynamic phases in a spring-block system , {\it Phys. Rev. E, 48}, 4779-4790,1993.

Johansen, A., D. Sornette, H. Wakita, U. Tsunogai, W.I. Newman and H. Saleur,
Discrete scaling in earthquake precursory phenomena : evidence in the Kobe earthquake,
Japan, {\it J.Phys.I France, 6}, 1391-1402, 1996.

Jones, L.M., Foreshocks, aftershocks and earthquake probabilities - Accounting
for the Landers earthquake, {\it Bull. Seism. Soc. Am., 84}, 892-899, 1994.
   
Jones, R., S. \"Oberg, M.I. Heggie and P. Tole, {\it Ab initio} calculation of the
structure of molecular water in quartz, {\it Philos. Mag. Lett., 66}, 61-66, 1992.

Jones,  L.M., Foreshocks, aftershocks, and earthquake probabilities :
accounting for the Landers earthquake, {\it Bull. Seismological Soc. Am., 84},
892-899, 1994.

Jordan, T.H., Is the study of earthquakes a basic science? {\it Seismol. Res. Lett., 68 (2)},
259-261, 1997.

Kanamori, H., Mechanics of earthquakes, {\it Ann. Rev. Earth Planet. Sci. 22}, 207-237,
1994.

Kanamori, H., and D.L. Anderson, Theoretical basis of some empirical relations
in seismology, {\it Bull. Seism. Soc. Am., 65}, 1073-1095, 1975.

Kanamori, H., and C.R. Allen, Earthquake repeat time and average stress drop,
In {\it Earthquake source mechanics, Geophys. Monogr.}, ed. S. Das, J. Boatwright,
C.H. Scholz, Washington, DC\,: Am. Geophys. Union, 227-2235, 1986.

Kanamori, H., J. Mori, E. Hauksson, T.H. Heaton, L.K. Hutton and L.M. Jones,
Determination of earthquake energy release and $M_L$ using Terrascope, {\it Bull.
Seism. Soc. Am., 83}, 330-346, 1993.

Kagan, Y.Y., Observational evidence for earthquakes as a nonlinear dynamic process,
{\it Physica D 77}, 160-192, 1994.

Kagan, Y.Y., Seismic moment-frequency relation for shallow earthquakes: regional
comparison, {\it J. Geophys. Res., 102}, 2835-2852, 1997.

Kagan, Y.Y., Is earthquake seismology hard, quantitative science? preprint 1998
http://scec.ess.ucla.edu/ykagan.html

Keilis-Borok, V.I., editor, Intermediate-term earthquake prediction\,:
models, algorithms, worldwide tests, {\it Phys. Earth and Planet. Interiors, 61},
Nos. 1-2, 1990.

King, G.C.P., The accommodation of large strains in the upper lithosphere of
the Earth and other solids by self-similar fault systems; the
        geometrical origin of b-value, {\it Pure Appl. Geophys., 121}, 761-815, 1983.

King, G.C.P., Speculations on the geometry of the initiation and
termination processes of earthquake rupture and its relation to
morphology and geological structure, {\it Pure Appl. Geophys., 124}, 567-585, 1986.

King, G.C.P., R.S. Stein and J.B. Rundle, The growth of geological structures by
repeated earthquakes. 1. Conceptual framework, {\it J. Geophys. Research, 93},
13307-13318, 1988.

King, G.C.P., G. Bailey and D. Sturdy, Active tectonics and human
survival strategies, {\it J. Geophys. Res., 99}, 20063-20078, 1994.

Kirby, S.H., Introduction and digest to the special issue on chemical effects of
water on the deformation and strength of rocks, {\it J. Geophys. Res., 89},
3991-3995, 1984.

Kirby, S.H., Localized polymorphic phase transformations in
high-pressure faults and applications to the physical mechanism of deep
earthquakes, {\it J. Geophys. Res., 92}, 13789-13800, 1987.

Klein, W, Rundle, J.B. and C.D. Ferguson,
Scaling and nucleation in models of earthquake faults,
{\it Phys. Rev. Lett. 78}, 3793-3796, 1997.

Knipe, R.J., The interaction of deformation and metamorphism in slates, 
{\it Tectonophysics, 78}, 249-272, 1981.

Knipe, R.J. and G.E. Lloyd, Microstructural analysis of faulting in quartzite,
Assynt, NW Scotland\,: implications for fault zone evolution, {\it Pure Appl. Geophys., 143},
229-254, 1994.

Knopoff, L., Energy release in earthquakes, {Geophys. J. Roy. Astron. Soc., 1},
44-52, 1958.
   
Knopoff, L., Q, {\it Rev. Geophys., 2}, 625-660, 1964.

Knopoff, L., A selective phenomenology of the seismicity of Southern California, 
{\it Proc. Nat. Acad. Sci. USA 93}, 3756-3763, 1996a.

Knopoff, L., The organization of seismicity on fault networks,
{\it Proc. Nat. Acad. Sci. USA 93}, 3830-3837, 1996b.

Knopoff, L., in Proceedings of Scale invariance and Beyond, eds. B. Dubrulle, F. Graner
\& D. Sornette, Springer, Heidelberg, 1997.

Knopoff, L., and M.J. Randall, The compensated linear-vector dipole\,: a possible 
mechanism for deep earthquakes, {\it J. Geophys. Res. 75}, 4957-4963, 1970.

Knopoff, L. and D.Sornette,
Earthquake death tolls, {\it J.Phys.I France, 5}, 1681-1688, 1995.
 
Knopoff, L., T. Levshina, V.I. Keilis-Borok and C. Mattoni, Increased long-range
intermediate-magnitude earthquake activity prior to strong earthquakes in
California, {\it J. Geophys. Res., 101}, 5779-5796, 1996.

Kronenberg, A. K., S.H. Kirby, S.H. Aines and G.R. Rossman, 
Solubility and diffusional uptake of hydrogen
in quartz at high water pressures; implications for hydrolytic weakening, {\it J.
Geophys. Res., 91}, 12723-12741, 1986. 

Kuge, K., and T. Lay, Systematic non-double-couple components
of earthquake mechanisms\,: the role of fault zone irregularity, {\it J. Geophys.
Res., 99}, 15457-15467, 1994.

Kuhlmann-Wilsdorf, D., {\it Phys.Rev. A, 140}, 1599, 1965.

Lachenbruch, A.H., Frictional heating, fluid pressure and the resistance of
fault motion, {\it J. Geophys. Res., 85}, 6097-6112, 1980.

Lachenbruch, A.H., and J.H. Sass, Heat flow and energetics of the San
Andreas fault zone, {\it J. Geophys. Res., 85}, 6185-6222, 1980.

Lachenbruch, A.H., and J.H. Sass, The stress heat-flow paradox and thermal results 
from Cajon Pass, {\it Geophys. Res. Lett. 15}, 981-984, 1988.

Lachenbruch, A.H., and J.H. Sass, Heat flow from Cajon Pass, fault strength
and tectonic implications, {\it J. Geophys. Res., 97},
4995-5015, 1992.

Lachenbruch, A.H., J.H. Sass, G.D. Clow and R. Weldon, Heat flow at
Cajon Pass, California, revisited, {\it J. Geophys. Res., 100},
2005-2012, 1995.

Langer, J.S., J.M. Carlson, C.R. Myers and B.E. Shaw, Slip complexity in 
dynamical models of earthquake faults, {\it Proc. Nat. Acad. Sci. USA
93}, 3825-3829, 1996.

Li, Y.G., K. Aki, D. Adams, and A. Hasemi, Seismic guided waves in the fault zone
of the Landers, California, earthquake of 1992, {\it J. Geophys. Res., 99},
11705-11722, 1994.

Linde, A.T., M.T. Gladwin, M.J.S. Johnston, R.L. Gwyther and R.G. Bilham, A slow
earthquake sequence on the San Andreas fault, {\it Nature, 383}, 65-68, 1996.

Lindh,  A.G., Earthquake prediction - The seismic cycle pursued, {\it Nature, 348}, 580-581, 1990.

Lockner, D.A., and J.D. Byerlee, How geometrical constraints contribute to the
weakness of mature faults, {\it Nature, 363}, 250-252, 1993.

Lomnitz, C., Nonlinear behavior of clays in the great Mexico earthquake, {\it C. R.
Acad. Sci. Paris II, 305}, 1239-1242, 1987. 

Lomnitz-Adler, J., Model for steady-state friction, 
{\it J. Geophys. Res., 96}, 6121-6131, 1991.

Lonsdale, K., The geometry of chemical reactions in single crystals, 
in Physics of the solid state, S. Balakrishna, M. Krishnamurthi and B.R. Rao, eds.,
Academic Press, London and New York, 43-61, 1969.

Lu, Z.W., and A. Zunger, Unequal wave vectors in short-range versus long-range
ordering in intermetallic compounds, {\it Phys. Rev. B, 50}, 6626-6636, 1994.

MacDonald, G.J.F., Orientation of anisotropic minerals in a stress field, 
{\it Geol. Soc. Am. Memoir, 79}, 1-8, 1957.

Madariaga, R., Dislocations and earthquakes, in Les Houches, Session XXXV, 1980
-- Physics of defects, R. Balian et al., eds, North Holland Publishing Company, 1981.

Maeda, K., The use of foreshocks in probabilistic prediction along the Japan and 
Kuril trenches, {\it Bull. Seism. Soc. Am., 86}, 242-254, 1996.

Main, I., Statistical physics, seismogenesis and seismic hazard, {\it Rev. Geophys., 34},
433-462, 1996.

Marais, S., V. Heine, C. Nex and E. Salje, Phenomena due to strain coupling in phase
transitions, {\it Phys. Rev. Lett., 66}, 2480-2483, 1991.

Martin, G., Transformation de phase et plasticit\'e, {\it Ann. Chim. Fr.,
6}, 46-58, 1981.

Martin, G., and P. Bellon, {\it Solid State Physics, 50}, 189-331, Ehrenreich and
Spaepen eds, Academic press, 1997.

Massonnet, D., K. Feigl, M. Rossi and F. Adragna, Radar interferometric mapping of
deformation in the year after the Landers earthquake, 
{\it Nature, 369}, 227-230, 1994.

Massonnet, D., W. Tatcher and H. Vadon, Detection of postseismic fault-zone
collapse following the Landers earthquake, {\it Nature, 382}, 612-614, 1996.

Maveyraud, C., W. Benz, G. Ouillon, A. Sornette and D. Sornette,
Solid friction at high sliding velocities\,: an explicit 3D dynamical
SPH approach, preprint 1998.

Melosh, H.J., Dynamical weakening of faults by acoustic fluidization,
{\it Nature, 379}, 601-606, 1996.

Miltenberger, P., D. Sornette and C. Vanneste,  
Fault self-organization as optimal random paths selected by critical spatio-temporal dynamics of
earthquakes, Phys.Rev.Lett. 71, 3604-3607, 1993.

Mogi, K., Some features of recent seismic activity in and near Japan
{2}: activity before and after great earthquakes, {\it Bull. Eq. Res. Inst. Tokyo
Univ., 47}, 395-417, 1969.

Moore, D.E., D.A. Lockner, R. Summers, M. Shengli et al., Strength of
chrysotile-serpentinite gouge under hydrothermal conditions - can it explain
a weak San Andreas fault? {\it Geology, 24}, 1041-1044, 1996.

Mora, P and D. Place, Simulation of the frictional stick-slip
instability, {\it Pure and Applied Geophysics, 143}, 61-87, 1994.

Mori, J., and H. Kanamori, Initial rupture of earthquakes in the 1995 Ridgecrest,
California sequence, {\it Geophys. Res. Lett., 23}, 2437-2440, 1996.

Morrow, C., B. Radney and J. Byerlee, Frictional strength and the effective
pressure law of Montmorillonite and Illite clays, in Fault mechanics and transport
properties of rocks, B. Evans and T.-F. Wong, eds., Academic Press, London, 69-88
1992.

Mulargia, F., and P. Gasperini, Evaluating the statistical validity beyond change
of `VAN' earthquake precursors, {\it Geophys. J. Int., 111}, 32-44, 1992.

Nabarro, F. R. N., Theory of dislocations,  Oxford Clarendon, reedited by
Dover, 1967, p688-691. 

NASDA-CON-960003, 1997, Abstracts of International Workshop on
Seismo-Electromagnetics '97.

National Research Council; The role of fluids in crustal processes, Studies in
geophysics, Geophysics study committed, Commission on Geosciences, Environment and
Ressources, National Academic Press, Washington D.C., 1990.

Nicolas, A., and J.P. Poirier, Crystalline plasticity and solid state flow in
metamorphic rocks, John Wiley \& sons, London, 1976.

Nicolaysen, L.O., and J. Ferguson, Cryptoexplosion structures, shock
deformation and siderophile concentration related to explosive venting of
fluids associated with alkaline ultramafic magmas, {\it Tectonophysics, 171},
303-335, 1990.

Nitsan, U., and T.J. Shankland, Optical properties and electronic structure of
mantle silicates, {\it Geophys. J. R. Astron. Soc., 45}, 59-87, 1976.

Nur, A., and J. Walder, Time-dependent hydraulics of the earth's crust, 
in The role of fluids in crustal processes, Studies in
geophysics, Geophysics study committed, Commission on Geosciences, Environment and
Ressources, National Research Council, National Academic Press, Washington D.C.,
113-127, 1990.

Od\'e, H., Faulting as a velocity discontinuity in plastic deformation, {\it The
Geological Society of America Memoir, 79}, A symposium on Rock Deformation, D.
Griggs and J. Handin, eds., 293-321, 1960.

Ogawa, M., Shear instability in a viscoelastic material as the cause
of deep focus earthquakes {\it J. Geophys. Res., 92}, 13801-13810, 1987.

Okay, A.L., X. Shutong and A.M.C. Seng\"or, Coesite from the Dabie Shan eclogites,
central China, {\it Eur. J. Mineral., 1}, 595-598, 1989.

O'Neil, J.R., and T.C. Hanks, Geochemical evidence for water-rock interaction along
the San-Andreas and Garlock faults in California, {\it J. Geophys. Res. 85}, 6286-6292, 1980.

Orowan, E., Mechanism of seismic faulting,  {\it The
Geological Society of America Memoir, 79}, A symposium on Rock Deformation, D.
Griggs and J. Handin, eds., 293-321, 1960.

Ortlepp, W.D., Note on fault-slip motion inferred from a study of
micro-cataclastic particles from an underground shear rupture, {\it Pure Appl. Geophys., 139},
677-695, 1992.

Ortlepp, W.D., {\it Rock Fracture and rockbursts -- an illustrative study} (S.A.
Inst. Min. and Metall. Johannesburg, 1997).

Ostwald, W., The formation and changes of solids, {\it Z.Phys. Chem. Frankfurt 22}, 289-330, 1897.

Ouillon, G., C. Castaing and D. Sornette,
Hierarchical scaling of faulting, {\it J. Geophys. Res. 101}, 5477-5487, 1996.

Paquet, J., P. Fran\c cois and A. Nedelec, Effect of partial melting on rock
deformation\,: experimental and natural evidences on rocks of granitic
compositions, {\it Tectonophysics, 78}, 323-345, 1981.

Paterson, M.S., The determination of hydroxyl by infrared absorption in quartz,
silicate glasses and similar materials, {\it Bull. Mineral., 105}, 20-29, 1982.

Paterson, M.S., The interaction of water with quartz and its influence in
dislocation flow - an overview, in Karato, S., and M. Toriumi, eds., Rheology of
solids and of the earth, Oxford University Press, London, 1990.

Peacock, S.M., Blueschist-facies metamorphism, shear heating, and P-T-T paths
in subduction shear zones, {\it J. Geophys. Res. 97}, 17693-17707, 1992.

Pearson, C.F., J. Beava, D.J. Darby, G.H. Blick and R.I. Walcott, Strain
distribution accross the Australian-Pacific plate boundary in the central
South Island, New Zealand, from 1992 GPS and earlier terrestrial observations,
{\it J. Geophys. Res. 100}, 22071-22081, 1995.

Pecher, A., Experimental decrepitation and re-equilibration of fluid inclusions in
synthetic quartz, {\it Tectonophysics, 78}, 567-583, 1981.

Persson, B.N.J., and E. Tosatti, eds., Physics of sliding friction, NATO ASI Series
(Kluwer Academic Publishers, Dordrecht, 1996).

Pietronero L., P. Tartaglia P., Y.C. and Zhang, Theoretical studies of
self-organized criticality, {\it Physica A 173}, 22-44, 1991.

Pinkston, J., L. Stern and S. Kirby, Hydrothermal reactions on artificial fault
surfaces in dunite\,: fibrous mineral growth, slickensides and temperature
sensitivity of reaction weakening, {\it Eos Trans. AGU, 68}, 405, 1987.

Pisarenko, D., and P. Mora, Velocity weakening in a dynamical model of friction,
{\it Pure and Applied Geophysics, 142}, 447-466, 1994. 

Poole, P. H., F. Sciortino, T. Grande, H. E. Stanley and C. A. Angell,
Effect of hydrogen bonds on the thermodynamic behavior of liquid
water, {\it Phys.  Rev. Letters, 73}, 1632-1635, 1994.

Poole, P. H., U. Essmann, F. Sciortino, and H. E. Stanley, Phase diagram for
amorphous solid water, {\it Phys. Rev. E, 48}, 4605-4610, 1993.

Poole, P. H., F. Sciortino, U. Essmann and H. E. Stanley, Spinodal of liquid
water, {\it Phys. Rev. E, 48}, 3799-3817, 1993.

Poole, P. H., F. Sciortino, U. Essmann and H. E. Stanley, Is there a re-entrant
spinodal in liquid water, {\it J. Phys. IV France, 3}, 171-182, 1993.

Poole, P. H., F. Sciortino, U. Essmann and H. E. Stanley, Phase behaviour of
metastable water, {\it Nature, 360}, 324-328, 1992.

Post, R.L. Jr, High-temperature creep of Mt. Burnet Dunite,
{\it Tectonophysics, 42}, 75-110, 1977.

Purton, J., R. Jones, M. Heggie, S. \"Oberg, and C.R.A. Catlow, LDF pseudopotential
calculations of the $\alpha$-quartz structure and hydrogarnet defect, {\it Phys.
Chem. Minerals, 18}, 389-392, 1992.

Raleigh, C.B., K. Sieh, L.R. Sykes and D.L. Anderson, Forecasting
Southern California Earthquakes, {\it Science, 217}, 1097-1104, 1982.

Ranalli, G., Rheology of the earth, Allen \& Unwin, Boston, 1987.

Rao, M., and S. Sengupta, Droplet fluctuations in the morphology and kinetics
of martensites, {\it Phys. Rev. Lett., 78}, 2168-2171, 1997.

Renard, F., P. Ortoleva and J.P. Gratier, 
Pressure solution in sandstones: influence of clays and dependence on
   temperature and stress, {\it Tectonophysics, 280}, 257-266, 1997.

Renard, F., J.P. Gratier, P. Ortoleva, E. Brosse, et al., 
Self-organization during reactive fluid flow in a porous medium,
{\it Geophys. Res. Lett., 25}, 385-388, 1998.

Rice, J.R., Fault stress states, pore pressure distributions and the weakness of
the San Andreas fault, in Fault mechanics and transport properties in rocks (the
Brace volume), ed. Evans, B., and T.-F. Wong, Academic, London, 475-503, 1992.

Rice, J.R., Spatio-temporal complexity of slip on a fault, {\it J.Geophys.Res. 98}, 9885-9907, 1993.

Robert, F., A.-M. Boullier and K. Firdaous, Gold-quartz veins in metamorphic terranes
and their bearing on the role of fluids in faulting, {\it J. Geophys. Res., 100},
12861-12879, 1995.

Rubie, D.C., Reaction-enhanced ductility\,: the role of solid-solid univariant
reactions in deformation of the crust and mantle, {\it Tectonophysics, 96},
331-352, 1983.

Ruina, A., Slip instability and state variable friction laws,
{\it J. Geophys. Res., 88}, 10359-10370, 1983.

Rundle, J.B., and W. Klein, New ideas about the physics of earthquakes,
{\it Review of Geophysics 33}, 283-286,1995.

Russo, P., R. Chirone, L. Massimilla and S. Russo, The influence of the frequency of
acoustic waves on sound-assisted fluidization of beds of fine particles, {\it Powder
Technology, 82}, 219-230, 1995.

Rydelek, P.A., I.S. Sacks and R. Scarpa, On tidal triggering of earthquakes at Campi
Flegrei, Italy, {\it Geophys. J. Int., 109}, 125-137, 1992.

Saleur, H., C.G. Sammis and D. Sornette, Discrete scale invariance,
complex fractal dimensions and log-periodic corrections in earthquakes,
{\it J.Geophys.Res., 101}, 17661-17677, 1996.

Salje, E. K. H., Phase transitions in ferroelastic and co-elastic crystals\,: an
introduction for mineralogists, material scientists, and physicists,
Cambridge, England; New York : Cambridge University Press, 1990.

Salje, E. K. H., Strain-related transformation twinning in minerals, {\it Neues
Jahrbuch fur Mineralogie-Abhandlungen, 163}, 43-86, 1991.

Salje, E. K. H., Application of Landau theory for the analysis of phase transitions
in minerals, {\it Phys. Rep., 215}, 49-99, 1992.

Sass, J.H., A.H. Lachenbruch, T.H. Moses and P. Morgan, Heat flow from 
a scientific research well at Cajon Pass, California, 
{\it J. Geophys. Res., 97}, 5017-5030, 1992.

Schallamach, A., How does rubber slide? {\it Wear 17}, 301-312, 1971.

Scholz, C.H., Shear heating and the state of stress on faults, {\it J. Geophys. Res. 85},
6174-6184, 1980.

Scholz, C.H., The mechanics of earthquakes and faulting, Cambridge University
Press, Cambridge, 1990.

Scholz C.H., Earthquakes and Faulting: self-organized critical phenomena with
characteristic dimension.; in Triste and D. Sherrington (eds), Spontaneous
formation of space-time structures and criticality,Kluwer Academic Publishers,
Netherlands, 41-56, 1991.

Scholz, C.H., Paradigms or small change in earthquake mechanics, in Fault mechanics
and transport properties in rocks (the Brace volume), ed. Evans, B., and T.-F.
Wong, Academic, London, 505-517, 1992a.

Scholz, C.H., Weakness amidst strength, {\it Nature, 359}, 677-678, 1992b.

Scholz, C.H., Whatever happened to earthquake prediction? {\it Geotimes, 42}, 16-19, 1997.

Scholz, C.H., Earthquakes and friction laws, {\it Nature, 391}, N6662, 37-42, 1998.

Scholz, C.H., C.A. Aviles and S.G. Wesnousky, Scaling differences between large
interplate and intraplate earthquakes, {\it Bull. Seism. Soc. Am. 76}, 65-70, 1986.

Schmittbuhl, J., J.P. Vilotte and S. Roux, Dynamic friction of self-affine
surfaces, {\it J. Phys. II France, 4}, 225-237, 1994.

Schmittbuhl, J., J.P. Vilotte and S. Roux, Velocity weakening friction -- A
renormalization approach, {\it J. Geophys. Res., 101}, 13911-13917, 1996.

Sciortino, F.,  P. H. Poole, H. E. Stanley and S. Havlin, Lifetime of the bond
network and gel-like anomalies in supercooled water, {\it Phys. Rev. Lett. 64},  
1686-1689, 1990.

Scott, D., Seismicity and stress rotation in a granular model of 
the brittle crust, {\it Nature, 381}, 592-595, 1996.

Scott, D.R., C.J. Marone and C.G. Sammis, The apparent friction of granular fault
gouge in sheared layers, {\it J. Geophys.Res., 99}, 7231-7246, 1994.

Segall, P., and J.R. Rice, Dilatancy, compaction and slip instability of a
fluid-infiltrated fault, {\it J. Geophys. Res., 100}, 22155-22171, 1995.

Shaw, B.E., Generalized Omori law for aftershocks and foreshocks from a simple
dynamics, {\it Geophys. Res. Lett., 20}, 907-910, 1993.

Shaw, B.E., Complexity in a spatially uniform continuum fault model,
{\it Geophys. Res. Lett., 21}, 1983-1986, 1994.

Shaw, B.E., Frictional weakening and slip complexity in earthquake faults, 
{\it J. Geophys. Res., 102}, 18239-18251, 1995.

Shaw, B.E., Model quakes in the two-dimensional wave equation, 
{\it J. Geophys. Res., 100}, 27367-27377, 1997.

Shen, Z.-K., D.D. Jackson, Y. Feng, M. Cline, M. Kim, P.
Fang and Y. Bock, Postseismic deformation following the Landers earthquake,
California, 28 June 1992, {\it Bull. Seism. Soc. Am., 84}, 780-791, 1994.

Shen, Z.-K., D.D. Jackson and B.X. Ge, Crustal deformation across and beyond the
Los Angeles basin from geodetic measurements, 
{\it J. Geophys.Res., 101}, 27957-27980, 1996.

Shnirman, M.G.,  and E.M. Blanter, Self-organized criticality in a mixed hierarchical system,
preprint 1998.

Sibson, R.H., Interactions between temperature and pore fluid pressure during earthquake
faulting and a mechanism for partial or total stress relief, {\it Nature, 243}, 66-68, 1973.

Sibson, R.H., Power dissipation and stress levels on faults in the upper crust,
{\it J. Geophys. Res. 85}, 6239-6247, 1980.

Sibson, R.H., Fluid flow accompanying faulting\,: field evidence and models, in
Earthquake prediction, an international review, Amer. Geophys. Union, D.W.
Simpsonand P.G. Richards, eds., {\it Maurice Ewing Ser., 4}, 593-603, 1982.

Sibson, R.H., An assessment of field evidence for `Byerlee' friction, {\it
Pure Appl. Geophys., 142}, 645-662, 1994.

Sibson, R.H., J. McM. Moore and A.H. Rankin, Seismic pumping - a hydrothermal
fluid transport mechanism, {\it J. Geol. Soc. Lond., 131}, 652-659, 1975.

Silver, P.G., S.L. Susan, T.C. Wallace, C. Meade, S.C. Myers, D.E. James and R.
Kuehnel, Rupture characteristics of the deep Bolivian earthquake of 9 June 1994
and the mechanism of deep-focus earthquakes, {\it Science, 268}, 69-73, 1995.

Sleep, N.H., and M.L. Blanpied, Creep, compaction and the weak rheology of major
faults, {\it Nature, 359}, 687-692, 1992.

Smirnov, M.B., and A.P. Mirgorodsky, Lattice dynamical study of the $\alpha-\beta$
phase transition of quartz\,: soft-mode behavior and elastic anomalies, {\it Phys.
Rev. Lett., 78}, 2413-2416, 1997.

Smith, D.C., and M.A. Lappin, Coesite in the Straumen kyanite-eclogite pod, Norway,
{\it Terra Research, 1}, 47-56, 1989.

Smith, D.L., and B. Evans, Diffusional crack healing in quartz, 
{\it J. Geophys.Res., 89}, 4125-4135, 1984.

Snay, R.A., M.W. Cline, C.R. Philipp, D.D. Jackson et al., Crustal velocity field
near the big bend of California San Andreas fault,
{\it J. Geophys.Res., 101}, 3173-3185, 1996.

Snieder, R. and T. van Eck, Earthquake prediction: a political problem? 
{\it Geologische Rundschau, 86}, 446-463, 1997.

Sokoloff, J.B., Theory of dynamical friction between idealized sliding surfaces, 
{\it Surf. Sci., 144}, 267-272, 1984.

Sornette D., Self-organized criticality in plate tectonics, in the proceedings of
the NATO ASI "Spontaneous formation of space-time structures and criticality",
Geilo, Norway 2-12 april 1991, edited by T. Riste and D. Sherrington, Kluwer
Academic Press, p.57-106, 1991.

Sornette, A., and Sornette D., Self-Organized criticality and earthquakes, 
{\it Europhys. Lett. 9}, 197-202, 1989.

Sornette, A., P. Davy and D. Sornette, Growth of fractal fault patterns,
{\it Phys.Rev.lett. 65}, 2266-2269, 1990.

Sornette, D., Davy P. and Sornette A., Structuration of the lithosphere 
in plate tectonics as a Self-Organized Critical phenomena. {\it J. Geophys. Res. 95}, 
17353-17361, 1990.

Sornette, D., C. Vanneste and L. Knopoff, Statistical model of earthquake foreshocks, 
{\it Phys. Rev. A, 45}, 8351-8357, 1992.

Sornette, D., P. Miltenberger and C. Vanneste, 
Statistical physics of fault patterns self-organized by repeated earthquakes, {\it Pure
Appl. Geophys. 142},
491-527, 1994a.

Sornette, A.,  D. Sornette and P. Evesque, 
Frustration and disorder in granular media and tectonic blocks : implications for earthquake
complexity, {\it Nonlinear Processes in Geophysics 1}, 209-218, 1994b.

Sornette, D., and C.G. Sammis, Complex critical exponents from renormalization group
theory of earthquakes\,:  Implications for earthquake predictions, {\it J.Phys.I
France, 5}, 607-619, 1995.

Sornette, D., P. Miltenberger and C. Vanneste, Statistical physics of
fault patterns self-organized by repeated earthquakes : synchronization versus self-organized
criticality,  in Recent Progresses in Statistical Mechanics and Quantum Field Theory, 
Proceedings of the conference Statistical Mechanics and Quantum Field Theory, 
 eds. P. Bouwknegt, P. Fendley, J. Minahan, D. Nemeschansky, K. Pilch,
H. Saleur and N. Warner, World Scientific, Singapore, 1995, pp .313-332. 

Sornette, D., and C. Vanneste, Fault self-organization by repeated earthquakes
 in a quasi-static antiplane crack model
{\it Nonlinear Processes in Geophysics 3}, 1-12, 1996.

Stel, H., Crustal growth in cataclasites\,: diagnostic microstructures and
implications, {\it Tectonophysics, 78}, 585-600, 1981.

Streit, J.E., Low frictional strength of upper crustal faults: A model,
{\it J. Geophys. Res., 102}, 24619-24626, 1997.

Sturtevant, B., H. Kanamori and E.E. Brodsky, Seismic triggering by
rectified diffusion in geothermal systems, {\it J. Geophys. Res. 101},
25269-25282, 1996.

Sykes, L.R., and S. Jaum\'e, Seismic activity on neighbouring faults
as a long-term precursor to large earthquakes in the San Francisco Bay Area,
{\it Nature, 348}, 595-599, 1990.

Tanguy, A., and P. Nozi\`eres, First-order bifurcation landscape in a 2D geometry -
The example of solid friction, {\it J. Phys. I (France) 6}, 1251-1270, 1996.

Tanguy, A and S. Roux, Memory effects in friction over correlated surfaces, 
{\it Phys. Rev. E, 55}, 2166-2173, 1997.

Thiel, M., A. Willibald, P. Evers, A. Levchenko, P. Leiderer and S. Balibar,
Stress-induced melting and surface instability of $^4He$ crystals, {\it Europhys.
Lett., 20}, 707-713, 1992.

Thurber, C., S. Roecker, W. Ellsworth, Y. Chen, W. Lutter and R. Sessions, 
Two-dimensional seismic image of the San Andreas fault in the northern Gabilan
range, central California\,: evidence for fluids in the fault zone,
{\it Geophys. Res. Lett. 24}, 1591-1594, 1997.

Tsunogai, U., and  Wakita H., Precursory chemical changes in ground water -
 Kobe earthquake, Japan, {\it Science, 269}, 61-63, 1995.
 
Tsunogai, U., and  Wakita H.,  Anomalous changes in groundwater chemistry - 
 Possible precursors of the 1995 Hyogo-ken Nanbu earthquake, Japan, 
 {\it J. Phys. Earth, 44}, 381-390, 1996.

Tsuruoka, H., M. Ohtake and H. Sato, Statistical test of the tidal triggering of
earthquakes: contribution of the ocean tide loading effect, {\it Geophys. J. Int., 122},
183-194, 1995.

Tsutsumi, A., and T. Shimamoto, Frictional properties of monzodiorite and gabbro during
seismogenic fault motion, {\it J. Geol. Soc. Japan, 102}, 240-248, 1996.

Tsutsumi, A., and T. Shimamoto, High-velocity frictional properties of gabbro,
{\it Geophys. Res. Lett., 24}, 699-702,1997.

Tullis, J.A., Prefered orientation in experimentally deformed quartzites, {\it
Ph.D. Thesis, Univ. California, Los Angeles}, 344 pp, 1971.

Tullis, J., J.M. Christie and D.T. Griggs, Microstructures and preferred
orientations of experimentally deformed quartzites, {\it Geol. Soc. Am. Bull.,
84}, 297-314, 1973.

Turcotte, D.L., and G. Schubert, Geodynamics, applications
of continuum physics to geological problems, John Wiley and Sons, New York, 1982.

Turner, J.A. and R.L. Weaver, Diffuse energy propagation on heterogeneous plates -
Structural acoustics radiative transfer theory, {\it J. Acoust. Soc. Am., 100},
3686-3695, 1996.

Van Albada, M.P., B.A. Van Tiggelen, A. Lagendijk and
A. Tip, Speed of propagation of classical waves in strongly scattering media,
{\it Phys. Rev. Lett., 66}, 3132-3135, 1991.

Van Tiggelen, B.A., and A. Lagendijk, Rigorous treatment of the speed of
diffusing classical waves, {\it Europhys. Lett. ,  23}, 311-316, 1993.

Varotsos, P., and M. Lazaridou, Latest aspects of earthquake prediction in Greece
based on seismic electric signals, {\it Tectonophysics, 188}, 321-347, 1991.

Vernon, R.H., Optical microstructure of partly recrystallized calcite in some
naturally deformed marbles, {\it Tectonophysics, 78}, 601-612, 1981.

Vidale, J.E., D.C. Agnew, M.J.S. Johnston and D.H. Oppenheimer, Absence of earthquake
correlation with earth tides; an indication of high preseismic fault stress rate, 
{\it J. Geophys. Res.} in press, 1998.

Volmer, A., and T. Nattermann, Towards a statistical theory of solid dry friction
http://xxx.lanl.gov/cond-mat/9612206

Walcott R.I. et al., Geodetic strains and large earthquakes in the axial tectonic
belt of North Island, New Zealand, {\it J. Geophys. Res. 83}, 4419-4429, 1978.

Walcott R.I. et al., Strain measurements and tectonics of New Zealand, 
{\it Tectonophysics , 52}, 479, 1979.

Walmann, T., A. Malthe-Sorenssen, J. Feder, T. Jossang, P. Meakin and H.H. Hardy,
Scaling relations for the lengths and widths of fractures, {\it Phys. Tev. Lett.,
77}, 5393--5396, 1996.

Walther, J.V., Fluid dynamics during progressive regional metamorphism, in
The role of fluids in crustal processes, Studies in
geophysics, Geophysics study committed, Commission on Geosciences, Environment and
Ressources, National Research Council, National Academic Press, Washington D.C.,
64-71, 1990.

Wang, X., and J.G. Liou, Coesite-bearing eclogite from the Dabie mountains in central
China, {\it Geology, 17}, 1085-1088, 1989.

Wallace, R.E., ed., U.S. Geological Survey, Professional Paper 1515, ``The San Andreas Fault System,
California'', published by United States Government Printing
Office, Washington in 1990 and 1991 (the Books and Open File
Reports Section, U.S. Geological Survey, Federal Center, Box 25425, Denver, CO, 80255).

Weber, K., Kinematic and metamorphic aspects of cleavage formation in very
low-grade metamorphic slates, {\it Tectonophysics, 78}, 291-306, 1981.

Westwood, A.R.C., J.S. Ahearn and J.J. Mills, Developments in the theory and
applications of chemomechanical effects, {\it Colloids and Surfaces, 2}, 1-35, 1981.

Westwood, A.R.C., and J.S. Ahearn, Adsorption-sensitive flow and fracture of solids,
in {\it Physical chemistry of the solid state\,: applications to metals and their
compounds}, ed. F. Lacombe (Elsevier Science Publishers B.V., Amsterdam),
65-85, 1984.

Wintsch, R.P., R. Christoffersen and A.K. Kronenberg, Fluid-rock reaction
weakening of fault zones, {\it J. Geophys. Res., 100},13021-13032, 1995. 

Wolverton, C., and A. Zunger, $Ni-Au$\,: testing ground for theories of phase
stability, cond-mat/9701203, preprint 1997.
 
Wyss, M., ed., Evaluation of proposed earthquake precursors, American
Geophysical Union, Washington, DC, 1991.
     
Wyss, M., Second round of evaluations of proposed earthquake precursors,
{\it Pure Appl. Geophys. 149}, 3-16, 1997.

Wyss, M., R.L. Aceves and S.K. Park; 
Geller, R.J., D.D. Jackson, Y.Y. Kagan and P. Mulargia,
Cannot earthquakes be predicted? {\it Science, 278}, 487-490, 1997.

Xu, H.-J., and L. Knopoff, Periodicity and chaos in a one-dimensional dynamical
model of earthquakes, {\it Phys. Rev. E, 50}, 3577-3581, 1994.

Zhang Y.-C., Scaling Theory of Self-organized Criticality, {\it Phys. Rev. Lett. 63},
470-473, 1989.

Zhao, D.P., and H. Kanamori, The 1994 Northridge earthquake -- 3-D crustal
structure in the rupture zone and its relation to the aftershock locations and
mechanism, {\it Geophys. Res. Lett., 22}, 763-766, 1995.

Zhao, X.-Z., R. Roy, K.A. Cherlan and A. Badzian, Hydrothermal growth of diamond
in metal$-C-H_2O$ systems, {\it Nature, 385}, 513-515, 1997.

Zoback, M.L., 1st-order and 2nd-order patterns of stress in the lithosphere - The
world stress map project, {\it J. Geophys. Res. 97}, 11703-11728, 1992a.

Zoback, M.L., Stress field constraints on intraplate seismicity in Eastern
North-America, {\it J. Geophys. Res. 97}, 11761-11782, 1992b.

Zoback, M.L., V. Zoback, J. Mount, J. Eaton, J. Healy et al., New evidence on the
state of stress of the San Andreas fault zone, {\it Science, 238}, 1105-1111, 1987.

}

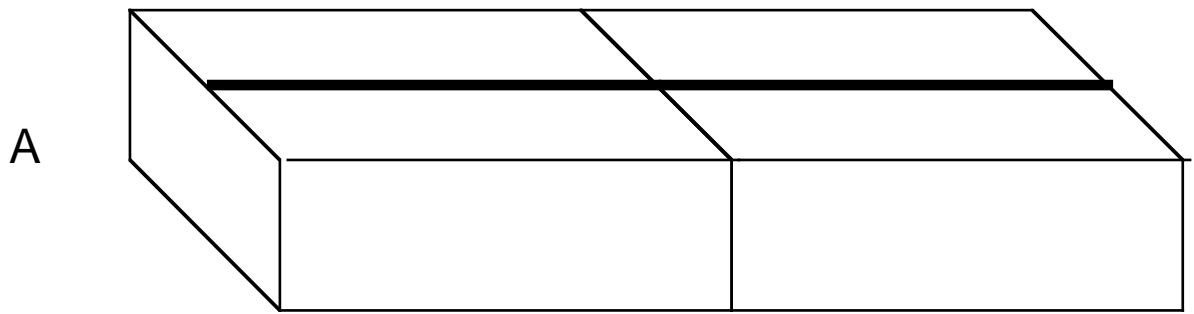
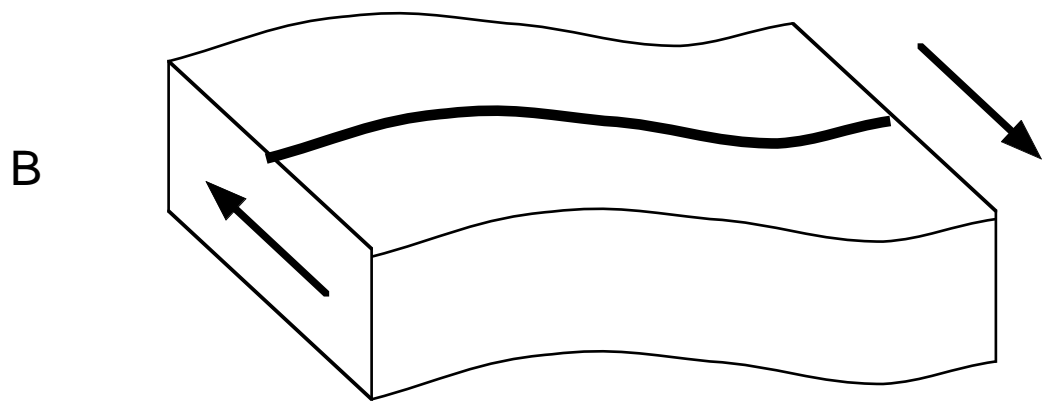
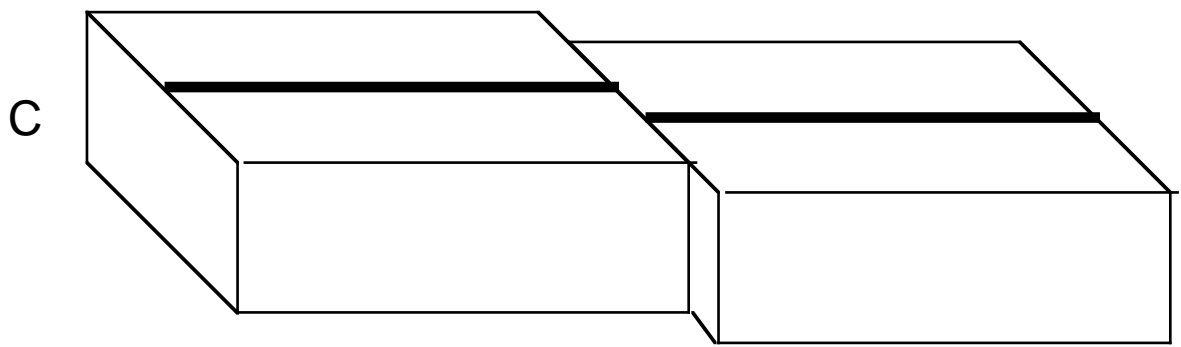
\end{document}